\documentclass{ws-ijmpd}
\usepackage{amsmath}
\newcommand{\overbar}[1]{\mkern 1.5mu\overline{\mkern-1.5mu#1\mkern-1.5mu}
\mkern 1.5mu}
\DeclareMathOperator{\sn}{sn}
\DeclareMathOperator{\sd}{sd}
\DeclareMathOperator{\K}{K}
\usepackage[colorlinks]{hyperref}
\usepackage[super,compress]{cite}
\begin{document}

\markboth{O.Chaschina,A.Sen,Z.Silagadze}
{On deformations of classical mechanics}

%
\catchline{}{}{}{}{}
%

\title{ON DEFORMATIONS OF CLASSICAL MECHANICS DUE TO PLANK-SCALE PHYSICS  }

\author{Olga I. Chashchina}
\address{\'{E}cole Polytechnique, Palaiseau, France\\
chashchina.olga@gmail.com}

\author{Abhijit Sen}
\address{Novosibirsk State University, Novosibirsk 630
090, Russia\\
abhijit913@gmail.com}

\author{Zurab K.~Silagadze}
\address{Budker Institute of Nuclear Physics and Novosibirsk State 
University, Novosibirsk 630 090, Russia\\
Z.K.Silagadze@inp.nsk.su}
\thispagestyle{plain}

\maketitle

\begin{history}
\received{Day Month Year}
\revised{Day Month Year}
\end{history}

\begin{abstract}
Several quantum gravity and string theory thought experiments indicate
that the Heisenberg uncertainty relations get modified at the Planck scale so 
that a minimal length do arises. This modification may imply a modification
of the canonical commutation relations and hence quantum mechanics at  
the Planck scale. The corresponding modification of classical mechanics is
usually considered by replacing modified quantum commutators by Poisson
brackets suitably modified in such a way that they retain their main properties
(antisymmetry, linearity, Leibniz rule and Jacobi identity). We indicate that
there exists an alternative interesting possibility. Koopman-Von Neumann's 
Hilbert space formulation of classical mechanics allows, as Sudarshan
remarked, to consider the classical mechanics as a hidden variable quantum 
system. Then the Planck scale modification of this quantum system naturally
induces the corresponding modification of dynamics in the classical substrate.
Interestingly, it seems this induced modification in fact destroys the
classicality: classical position and momentum operators cease to be commuting
and hidden variables do appear in their evolution equations. 
\end{abstract}



\section{Introduction}
Quantum mechanics is one of the most successful theories in science. At
present no single experimental fact indicates its breakdown. On the contrary,
we have every reason to believe that quantum mechanics encompasses every
natural phenomena. Nevertheless some spell of mystery still accompanies
quantum mechanics. Richard Feynman, worlds one of the best experts in quantum
mechanics, expressed this feeling most eloquently \cite{1-1}:

``We always have had (secret, secret, close the doors!) we always have had 
a  great deal of difficulty in understanding the world view that quantum 
mechanics  represents. At least I  do, because I'm an old enough man that 
I  haven't got to the point that this stuff is obvious to me. Okay, I  still 
get nervous with it. And therefore, some of the younger students \ldots  
you  know  how it always is, every new idea, it takes a  generation or two 
until it becomes obvious that there's no real problem. It has not yet become 
obvious to me that there's no real problem. I  cannot define the real problem, 
therefore I  suspect there's no real problem, but  I'm note sure there's no 
real problem.'' 

Superposition principle, inherent of quantum mechanics in which states
of quantum systems evolve according to linear Schr\"{o}dinger equation,
maybe is the core reason of our uneasiness with quantum mechanics. If 
classical mechanics is considered as a limit of quantum mechanics then the
superposition principle must hold in classical mechanics too \cite{1-2}.
However in the classical world, as it is revealed to us by our perceptions,
we never experience Schr\"{o}dinger cat states (except perhaps in art, see 
\cite{1-3}) and a widespread belief is that the environment induced decoherence
explains why ( see \cite{1-4,1-5} and references therein. For a contrary 
view, however see \cite{1-6,1-7}).  

A particularly striking example of decoherence is chaotic rotational motion 
of Saturn's potato-shaped moon Hyperion. The orbit of Hyperion around Saturn 
is fairly predictable, but the moon tumbles unpredictably as it orbits 
because its rotational motion is chaotic. It was argued \cite{1-8,1-9} that
if Hyperion were isolated from the rest of the universe, it would evolve
into a macroscopic Schr\"{o}dinger cat state of undefined orientation in 
a time period of about 20-30 years. However this never happens. Hyperion
is not isolated but constantly bombarded  by photons from the rest of the
universe causing its quantum state to collapse into a state of definite
orientation.

Gravitational interaction cannot be shielded. Therefore any object in the
universe is constantly bombarded by gravitons destroying macroscopic 
Schr\"{o}\-din\-ger cat states. It is expected, therefore, that gravity plays
a prominent role in the emergence of classicality \cite{1-10,1-11}.

However it can be argued  that the Planck scale should be viewed as 
a fundamental boundary of validity of the classical concept of spacetime, 
beyond which quantum effects cannot be neglected \cite{1-12}. A legitimate
question then is how this expected modifications of quantum mechanics and/or
gravity at the Planck scale influence the emergence of classicality. In this
paper we attempt to discuss some aspects of this question in the framework of
Koopman-von Neumann theory.

The manuscript is organized as follows. In the second section a brief 
over\-view of the generalized uncertainty principle is given. The third 
section provides fundamentals of the Koopman-von Neumann formulation of 
classical mechanics. In the fourth and fifth sections we describe 
modifications of classical mechanics expected then combining Koopman-von 
Neumann-Sudarshan perspective on classical mechanics with the generalized 
uncertainty principle. In the last section some concluding remarks are given.

\section{Generalized Uncertainty Principle}
Heisenberg's uncertainty relation, that implies the non-commutativity of the 
quantum mechanical observables, underlines the essential difference between 
classical and quantum mechanics \cite{2-1}. Analyzing his now famous thought 
experiment of measuring the position of an electron using a gamma-ray 
microscope, Heisenberg arrived at the conclusion that ``the more precisely 
is the position determined, the less precisely is the momentum known, and 
vice versa'' \cite{2-2}:
\begin{equation}
\delta q\,\delta p\sim\hbar.
\label{eq2-1}
\end{equation}
Here $\delta q$ is the uncertainty in the determination of the position 
of the electron $q$, and $\delta p$ is the perturbation in its momentum $p$,
canonically conjugate to $q$, induced by the measurement process.

The precise meaning of ``uncertainty'' was not defined in Heisenberg's
paper who used heuristic arguments and some plausible measures of 
inaccuracies in the measurement of a physical quantity and quantified them 
only on a case-to-case basis as ``something like the mean error'' \cite{2-3}. 
After publication of \cite{2-2}, ``which gives an incisive analysis of the 
physics of the uncertainty principle but contains little mathematical 
precision'' \cite{2-4}, attempts to overcome its mathematical deficiencies 
were soon undertaken by Kennard \cite{2-5} and Weyl \cite{2-6}. They proved
the inequality, valid for any quantum state,
\begin{equation}
(\Delta q)^2\,(\Delta p)^2\ge\frac{\hbar^2}{4},
\label{eq2-2}
\end{equation}
where $(\Delta q)^2$ and $(\Delta p)^2$ are the variances (the second moment 
about the mean value) of $\hat q$ and  $\hat p$ defined as $(\Delta q)^2=
<\nolinebreak \hat q{\,^2}>-<\hat q>^2$ and similarly for $\hat p$. 
As usual, the mean value of a quantum-mechanical operator $\hat A$, in the 
quantum state $|\Psi>$, is defined as follows (we are considering a 
one-dimensional case, for simplicity)   
\begin{equation}
<\Psi|\hat A|\Psi|>=\int dq\,\Psi^*(q)(\hat A\Psi)(q).
\label{eq2-3}
\end{equation}
Taking the standard deviation $\Delta A$ (the square-root from the variance 
$(\Delta A)^2=\-<\hat A^{\,2}>-<\hat A>^2$) as a measure of 
indeterminacy (uncertainty) of the observable $\hat A$ in the quantum state 
$|\Psi>$ seems very natural from the point of view of the classical 
probability theory where the standard deviation is considered as a measure of 
fluctuations. Indeed, soon Ditchburn established the relation $\Delta q=
\delta q/\sqrt{2}$ between Heisenberg's $\delta q$ and Weyl-Kennard's 
$\Delta q$ and proved that  the equality in (\ref{eq2-2}) can be achieved for 
Gaussian probability distributions only \cite{2-7,2-8}.

From the mathematical point of view, we can put $<q>=0$ and $<p>=0$ in 
(\ref{eq2-2}) without loss of generality \cite{2-6}. Indeed, we can achieve
$<q>=0$ by suitable redefinition of the $q$-coordinate origin, and $<p>=0$
--- by multiplying the wave function by $\exp{(-i<p>\,q/\hbar)}$ without
changing the probability density associated with it. In this case 
(\ref{eq2-2}) becomes a mathematical statement about (normalized) 
square-integrable functions $\Psi(q)$:
\begin{eqnarray} &&
-\left(\int\limits_{-\infty}^\infty q^2\,|\Psi(q)|^2\,dq\right)
\left(\int\limits_{-\infty}^\infty \Psi^*(q)\,\frac{d^2\Psi(q)}{dq^2}\,dq
\right)=\nonumber \\ && 
\left(\int\limits_{-\infty}^\infty q^2\,|\Psi(q)|^2\,dq\right)
\left(\int\limits_{-\infty}^\infty \left|\frac{d\Psi}{dq}\right |^2\,dq
\right)\ge\frac{1}{4}.
\label{eq2-4}
\end{eqnarray}
The three-dimensional sibling of (\ref{eq2-4}), known as the Heisenberg 
inequality, is \cite{2-9}
\begin{equation}
\left(\int \mathbf{r}^2\,|\Psi(\mathbf{r})|^2\,d^3 r\right)
\left(\int\left|\boldsymbol{\nabla}\Psi(\mathbf{r})\right |^2\,d^3 r\
\right)\ge\frac{9}{4}.
\label{eq2-5}
\end{equation}
If an electron in the hydrogen atom  is localized around the origin, then
(\ref{eq2-5}) tells us that its momentum (and hence kinetic energy) will be
large. It is tempting to use this fact to get a lower bound on the electron's
energy in the hydrogen atom and and thus prove its stability. Although it is
a common practise to use such kind of reasoning for estimation of the hydrogen
atom size and its ground state energy \cite{2-10,2-11,2-12,2-13}, the truth
is that Heisenberg inequality is too weak to ensure the hydrogen 
atom stability or stability of matter in general \cite{2-14,2-15}. The reason
is that the first multiple in (\ref{eq2-5}) can be very large even in the case
when the main part of the wave function (its modulus squared) is localized 
around the origin, if only the remaining small part is localized very 
far away. 

However mathematically the uncertainty principle in the form of 
Eq.(\ref{eq2-5}) is just an expression of the fact from harmonic analysis that 
``A nonzero function and its Fourier transform cannot both be sharply 
localized'' \cite{2-4}. Heisenberg inequality (\ref{eq2-5}) is just one 
attempt to make the above given sloppy phrase mathematically precise. But, as 
we have seen above for hydrogen atom, from the physics perspective the  
standard deviation is not always an adequate measure of localization and 
quantum uncertainty \cite{2-16,2-17}. In such situations (in particular then 
considering the matter stability problem \cite{2-14,2-15}) other uncertainty 
principles prove to be more useful. The examples include Hardy uncertainty 
principle \cite{2-15,2-18}
\begin{equation}
\left(\int \frac{1}{\mathbf{r}^2}\,|\Psi(\mathbf{r})|^2\,d^3 r\right)^{-1}
\left(\int\left|\boldsymbol{\nabla}\Psi(\mathbf{r})\right |^2\,d^3 r\
\right)\ge\frac{1}{4},
\label{eq2-6}
\end{equation}
or Sobolev inequality \cite{2-14}
\begin{equation}
\left(\int |\Psi(\mathbf{r})|^6\,d^3 r\right)^{-1/3}
\left(\int\left|\boldsymbol{\nabla}\Psi(\mathbf{r})\right |^2\,d^3 r\
\right)\ge\frac{3}{4}\,(4\pi^2)^{2/3},
\label{eq2-7}
\end{equation}
which, in some sense, is weaker than the Hardy inequality (\ref{eq2-6})
\cite{2-15}.

From the physics side, the uncertainty principle is more than just inequalities
from harmonic analysis. We can envisage at least three   manifestations
of uncertainty  relations \cite{2-3,2-19}. First of all the uncertainty  
relations relate intrinsic spreads of two conjugate dynamical variables 
in a quantum state. However Heisenberg in his seminal work speaks about 
unavoidable disturbance that a measurement process exerts on a pair of 
conjugate dynamical variables. Therefore we can understand the uncertainty  
relation also as an assertion about a relation between inaccuracies in 
measurements of conjugate dynamical variables. Namely, the relation that 
connects either inaccuracy of a measurement of one dynamical variable and 
the ensuing disturbance in the probability distribution of the conjugated 
variable, or inaccuracies of a pair of conjugate dynamical variables in any 
joint measurements  of  these  quantities.Although conceptually distinct,
these three manifestations of uncertainty  relations are closely related
\cite{2-19}.

The first facet of the uncertainty principle can be formalized most easily
by using second-order central moments of two conjugated quantum observables
(but remember a caveat that  the standard deviation is not always an adequate 
measure of quantum uncertainty \cite{2-17}). For any pair of quantum 
observables $\hat A$ and $\hat B$ we have three independent second-order 
central moments of their joint quantum distributions in a quantum state 
$|\Psi>$:
\begin{eqnarray} &&
[\Delta(A)]^2=<\Psi|(\hat A -\bar A)^2|\Psi>=\overbar{A^2}-\bar A^{\,2},\;\;
[\Delta(B)]^2=<\Psi|(\hat B -\bar B)^2|\Psi>=\nonumber \\ &&
\overbar{B^2}-\bar B^{\,2},\;\;
\Delta(A,B)=\frac{1}{2}<\Psi|(\hat A -\bar A)(\hat B -\bar B)+
(\hat B -\bar B)(\hat A -\bar A)|\Psi>=\nonumber \\ &&
\frac{1}{2}\overbar{(AB+BA)}-\bar A\,\bar B,
\label{eq2-8}
\end{eqnarray}
where overbar denotes the mean value of the corresponding observable in the
state $|\Psi>$. We have
$$\hat A\hat B=\frac{1}{2}(\hat A\hat B+\hat B\hat A)+
\frac{1}{2}(\hat A\hat B-\hat B\hat A),$$
and the first Hermitian part in the r.h.s has a real mean value, while the
mean value of the second anti-Hermitian part  is purely imaginary. Then
the Schwarz inequality $|<\psi|\phi>|^2\le\; <\psi|\psi>\,<\phi|\phi>$,
if we take $|\psi>=\hat A|\Psi>$, $|\phi>=\hat B|\Psi>$, will give
$$\overbar{A^2}\,\overbar{B^2}\ge |\overbar{AB}|^2=\left(\frac{\overbar{AB+BA}}
{2}\right)^2+\left |\frac{\overbar{AB-BA}}{2}\right|^2,$$
because  $<\hat A\Psi|\hat A\Psi>=\overbar{A^2}$,  $<\hat B\Psi|\hat B\Psi>=
\overbar{B^2}$ and for Hermitian observables $\hat A$ and $\hat B$,
$<\hat A\Psi|\hat B\Psi>=<\Psi|\hat A\hat B|\Psi>=\overbar{AB}$.

If we replace $\hat A$ and $\hat B$ by operators $\hat A-\bar A$ and
$\hat B-\bar B$ in the above given reasoning, we end up with the 
Schr\"{o}dinger uncertainty relation \cite{2-20} 
\begin{equation}
[\Delta(A)]^2[\Delta(B)]^2\ge [\Delta(A,B)]^2+\left |\frac{\overbar{AB-BA}}{2}
\right|^2.
\label{eq2-9}
\end{equation}
Heisenberg uncertainty relation (\ref{eq2-2}) is obtained from this more
general uncertainty relation if we take $\hat A=\hat x$, $\hat B=\hat p$,
use the canonical commutation relations
\begin{equation}
[\hat x_i,\,\hat p_j]\equiv \hat x_i,\,\hat p_j-\hat p_j\,x_i=i\,\hbar\,
\delta_{ij},
\label{eq2-10}
\end{equation}
and assume that the covariance $\Delta(x,p)$ equals to zero. Although 
Schr\"{o}dinger uncertainty relation is more general and symmetric (it remains 
invariant under rotations in phase space) than the Heisenberg uncertainty 
relation, or its generalization due to Robertson to any pair of observables
with zero covariance
\cite{2-21}
\begin{equation}
[\Delta(A)]^2[\Delta(B)]^2\ge \left |\frac{\overbar{AB-BA}}{2}
\right|^2,
\label{eq2-11}
\end{equation}  
the Schr\"{o}dinger uncertainty relation is strangely ignored in almost
all quantum mechanics textbooks and its usefulness was appreciated only
after 50 years from its discovery in connection with the description of
squeezed states in quantum optics for which the covariance $\Delta(x,p)$
doesn't equal zero \cite{2-8}.

As we see the uncertainty relations are intimately related to canonical
commutation relations (\ref{eq2-10}). To our best knowledge, Gleb Wataghin
was the first \cite{2-22,2-23} who suggested that both the commutation 
relations (\ref{eq2-10}) and the uncertainty principle (\ref{eq2-1}) might be
modified at high relative impulses in such a way that 
\begin{equation}
\delta q\,\delta p\sim\hbar\,f(p),\;\;\;\lim_{p\to 0} f(p)=1,
\label{eq2-12}
\end{equation}
which can lead to the existence of a lower limit of a measurable lengths.
Below we give a brief outline of the modern developments of these kind of 
ideas.

If we don't care about the momentum, the uncertainty relation (\ref{eq2-2})
does not forbid us to prepare a quantum state with the arbitrary small 
position uncertainty. However, as argued in 1964 by Mead \cite{2-24}, things
change if we take into account effects of gravity.\footnote{The fact that 
gravity can influence the uncertainty relation (\ref{eq2-2}) was realized 
much earlier by Matvei Bronstein \cite{2-24Br}.} The crux of the Mead's
argument is that the gravitational interaction between the electron and
photon in  Heisenberg microscope is a source of additional uncertainty in
the electron's position. Almost identical arguments can be found in later
publications \cite{2-24A,2-24B}.

The gravitational field of a photon was obtained in \cite{2-25,2-26} by
boosting the Schwarzschild space-time up to the speed of light by taking
the limit $V\to c$, $m\to 0$ such that the quantity $p=mV(1-V^2/c^2)^{-1/2}$ 
is held constant. The resulting space-time, for the photon with momentum $p$ 
moving in the $z$-direction, has the metric 
\cite{2-27}
\begin{equation}
ds^2=-2\left (du\,dv-d\zeta\,d\bar\zeta\right )-4\,\frac{p\,G}{c^3}\ln{
\left (\frac{\zeta\,\bar\zeta}{r_0^2}\right )}\,\delta(u)\,du^2,
\label{eq2-13}
\end{equation}
where $G$ is Newton's constant, $u=(ct-z)/\sqrt{2}$ and $v=(ct+z)/\sqrt{2}$ 
are  retarded and advanced null coordinates (light-cone coordinates), while 
the complex coordinates $\zeta=(x+i\,y)/\sqrt{2}$ and $\bar\zeta=(x-i\,y)/
\sqrt{2}$ parametrize the spatial hyperplane orthogonal to the photon's  
velocity vector. The parameter $r_0$ in (\ref{eq2-13}) is an arbitrary constant
of the dimension of length which  does not effect observable quantities
\cite{2-27A}.

The metric (\ref{eq2-13}) describes an impulsive gravitational wave: the 
space-time remains flat everywhere except $u=0$ null hyperplane, where it 
develops a delta-function singularity. This gravitational shockwave moves 
with the photon and when it meets with the electron within the Heisenberg 
microscope two physical effects take place: the timelike geodesic of the 
electron experiences a discontinuous jump in the null coordinate $v$ and gets
refracted in the transverse direction \cite{2-29}.

There are various subtleties here. The very concept of photon with sharply
defined momentum (energy), existing at $t=-\infty$, is an idealization. In 
reality one should take into account that the photon is produced at a finite 
instant of time and the corresponding light packet has a finite Fourier
support \cite{2-28}. Besides, because (\ref{eq2-13}) describes a situation
when a cause (photon) and the effect (the corresponding gravitational 
shockwave) propagate with the same speed of light, it is not altogether clear 
the gravitational field is related to the photon or it arises solely in the 
process of emission \cite{2-29}. At last, to cope with the presence of 
ill-defined highly singular products of generalized functions in the geodesic 
deviation equation, precise calculation of the above mentioned physical 
effects of the  gravitational shockwave on the test particles geodesics, 
requires either a suitable regularization procedure \cite{2-30}, 
or making use of the Colombeau algebra of generalized functions \cite{2-31}.

Anyway, for our purposes we need only an order of magnitude estimate of the
additional uncertainty in Heisenberg microscope due to gravity. This was
done in \cite{2-32} with the result that the additional uncertainty in 
electron's position due to gravitational attraction of the photon is
\begin{equation}
\Delta x_G\approx \frac{Gp}{c^3}\approx l_P^2\,\frac{\Delta p}{\hbar}.
\label{eq2-14}
\end{equation}
The second step follows from the fact that the electron momentum uncertainty 
$\Delta p$ must be of order of the photon momentum $p$. Here
\begin{equation}
l_P=\sqrt{\frac{G\hbar}{c^3}}
\label{eq2-15}
\end{equation}
is the Planck length.

If we add this new uncertainty linearly to the original Heisenberg uncertainty
$\Delta x_H\approx \hbar/\Delta p$, we get the modified uncertainty principle
(the so called GUP --- generalized or gravitational uncertainty principle 
\cite{2-32})
\begin{equation}
\frac{\Delta x}{l_P}\approx \frac{\hbar}{l_P\,\Delta p}+
\frac{l_P\,\Delta p}{\hbar}.
\label{eq2-16}
\end{equation}
In this form the uncertainty principle is invariant under momentum inversion
$\frac{\hbar}{l_P\,\Delta p}\to \frac{l_P\,\Delta p}{\hbar}$. Another 
remarkable property of (\ref{eq2-16}) is that it predicts a minimum position
uncertainty $\Delta x_{min}=2\,l_P$ at a symmetric, with respect to the above 
mentioned momentum inversion, point $\Delta p=\hbar/l_P$.

Although the idea that a smallest length exists in nature can be traced back
to  Heisenberg and March \cite{2-33}, only relatively recent attempts to 
reconcile quantum mechanics with general relativity in string theory produced 
a solid foundation for it and for the generalized uncertainty principle
(see, for example, \cite{2-34,2-35} and review articles \cite{2-36,2-37}). 

As Robertson's version (\ref{eq2-11}) of the uncertainty principle shows,
the generalized uncertainty principle (\ref{eq2-16}) may imply a deformation
of the usual Heisenberg algebra of canonical commutation relations
(\ref{eq2-10}). Various versions of this deformation have been proposed
in the literature (see, for example, \cite{2-38} and references therein).

As was mentioned above, Wataghin was the first to suggest modification of
the canonical commutation relations at high energies. However, it was Snyder
who proposed a model of noncommutative spacetime, admitting a fundamental 
length but nevertheless being Lorentz invariant \cite{2-39}, non-relativistic 
version of which produces a concrete form of such  modification \cite{2-40}:
\begin{equation}
[\hat x_i,\,\hat x_j]=i\hbar\beta^\prime \hat J_{ij},\;\;
[\hat p_i,\,\hat p_j]=0,\;\;[\hat x_i,\,\hat p_j]=i\hbar 
(\delta_{ij}+\beta^\prime \hat p_i\hat p_j),
\;\; \hat J_{ij}=\hat x_i\hat p_j-\hat x_j\hat p_i,
\label{eq2-17}
\end{equation}
where $\beta^\prime$ is some constant, usually assumed to be of the order of 
$l_P^2/\hbar^2$, as (\ref{eq2-11}) and  (\ref{eq2-16}) relations do imply.

Snyder's work was ahead of its time and its importance was not immediately 
recognized. Meanwhile Mead \cite{2-24} and Karolyhazy \cite{2-42} investigated
uncertainties in measurements of space-time structure resulting from 
universally coupled  gravity and concluded that it is impossible  to  measure  
distances to a  precision  better  than Planck's  length. However very
few took seriously the idea that the Planck length could ever play 
a fundamental role in physics \cite{2-37,2-43}.  

The situation changed when developments in string theory revealed the very 
same impossibility of resolving distances smaller than Planck's  length,
and these developments inspired  Adler  and  Santiago's 1999 paper \cite{2-32}
that almost exactly reproduced Mead's earlier arguments \cite{2-37}. Various 
choices of deformed commutation relations  have been considered in the 
literature beginning from the Kempf et al. landmark paper \cite{2-44}. Let us
mention, for example, a version that generalizes the Snyder algebra 
(\ref{eq2-17}) \cite{2-45,2-46}:
\begin{eqnarray} &&
[\hat x_i,\,\hat x_j]=-i\hbar\left [(2\beta-\beta^\prime)+\beta(2\beta+
\beta^\prime)\hat p^2\right]\,\frac{\hat J_{ij}}{1+\beta\hat p^2},
\nonumber \\ &&
[\hat p_i,\,\hat p_j]=0,\;\;\;
\;\;[\hat x_i,\,\hat p_j]=i\hbar \left [(1+\beta \hat p^2)\delta_{ij}+
\beta^\prime \hat p_i\hat p_j\right ],
\label{eq2-18}
\end{eqnarray}
where $\beta$ is a new constant of the same magnitude as $\beta^\prime$,
$\hat p^2= \hat p_i \hat p^i$, and $\hat J_{ij}$ was defined in (\ref{eq2-17}).

A different type of modification of the canonical commutation relations 
was suggested by Saavedra and Utreras \cite{2-48}:
\begin{equation}
[\hat x_i,\,\hat p_j]=i\hbar \left (1+\frac{l_P}{c\hbar}\,H\right)
\delta_{ij}.
\label{eq2-19}
\end{equation}
One can say that in this case the configuration space becomes dynamical, much 
like the general relativity, because the commutation relations  (\ref{eq2-19})
depends on the system under study through the Hamiltonian $H$.

As we see, commutation relations (\ref{eq2-17}) and (\ref{eq2-18}) imply
a non-commutative spatial geometry. Mathematically this is a consequence of
Jacobi identity and our tacit assumption that components of momentum operators
do commute. Physical bases of this non-commutativity is a dynamical nature
of space-time in general relativity: it can be argued quite generally that
an  unavoidable change in the space-time metric when measurement processes 
involve energies of the order of the Planck scale destroys the commutativity 
of position operators \cite{2-47,2-47A,2-47B,2-47C}. 

There exists a vast and partly confusing literature on the modifications of
quantum mechanics and quantum field theory implied by the existence of
a minimal length scale (for a review and references see, for example,
\cite{2-36,2-37,2-49,2-50,2-51}).

\section{Koopman-von Neumann mechanics}
It is usually assumed that classical mechanics, in contrast to quantum 
mechanics, is a deterministic theory with the well defined trajectories
of underlying particles. However, if we realize the imperfect nature of 
classical measuring devices, which precludes the preparation of classical
systems with precisely known initial data, it becomes clear that ``the 
determinism of classical physics turns out to be an illusion, created by 
overrating mathematico-logical concepts. It is an idol, not an ideal in 
scientific research'' \cite{3-1}. Therefore, one can assume that 
a conceptually superior appropriate statistical description of classical 
mechanics is then given by Liouville equation (for simplicity, we consider 
a one-dimensional mechanical system with canonical variables $q$ and $p$)
\begin{equation}
i\frac{\partial \rho}{ \partial t}=\hat L \rho=i\{H,\rho\}=
i\left (\frac{\partial H}{ \partial q}\;\frac{\partial \rho}{ \partial p}-
\frac{\partial H}{ \partial p}\;\frac{\partial \rho}{ \partial q}\right ),
\label{eq3-1}
\end{equation}
which  gives a time-evolution of the phase-space probability density 
$\rho(q,p,t)$. Here $H$ is is the Hamiltonian and $\{,\}$ denotes the Poisson 
bracket. 

However, classical and quantum mechanics are different not only by 
inherently probabilistic nature of the latter. Mathematical structures
underlying these two disciplines are quite different. The mathematics 
underlying classical mechanics is a symplectic geometry of the phase space
\cite{3-2,3-3,3-4}, while quantum mechanics is based on the theory of Hilbert
spaces \cite{3-5}, rigged Hilbert spaces  \cite{3-6} or on their algebraic
counterpart --- the theory of $C^*$ algebras \cite{3-7}. In light of this 
difference in the underlying mathematical structure it is surprising that it 
is possible to give a Hilbert space formulation for classical mechanics too, 
as shown long ago in classic papers by Koopman \cite{3-8} and von Neumann 
\cite{3-9} (for modern presentation, see \cite{3-10} and references therein).
\footnote{Older references can be found in \cite{3-10A,3-10B}. It should be 
noted that, apparently independently from Koopman and von Neumann, and from 
each other, similar formalisms were suggested later by Sch\"{o}nberg 
\cite{3-10C} and by Della Riccia and Wiener \cite{3-10D}.} 

This translation of classical mechanics into the language of Hilbert spaces
is based on the crucial observation that, because the Liouville operator
\begin{equation}
\hat L=i\left (\frac{\partial H}{ \partial q}\;\frac{\partial}{ \partial p}-
\frac{\partial H}{ \partial p}\;\frac{\partial }{ \partial q}\right )
\label{eq3-2}
\end{equation}
is linear in derivatives, the square root of the probability density 
$\psi(q,p,t)=\sqrt{\rho(q,p,t)}$ obeys the same Liouville equation
(\ref{eq3-1}):
\begin{equation}
i\frac{\partial \psi(q,p,t)}{\partial t}=\hat L \psi(q,p,t).
\label{eq3-3}
\end{equation}
Moreover, if we assume that $\psi(q,p,t)$ in (\ref{eq3-3}) is a complex 
function $\psi(q,p,t)=\sqrt{\rho(q,p,t)}\,e^{iS(q,p,t)}$, then (\ref{eq3-3}) 
implies that the amplitude and phase evolve independently through the 
Liouville equations:
\begin{equation}
i\frac{\partial \sqrt{\rho}}{\partial t}=\hat L \sqrt{\rho},\;\;\;
i\frac{\partial S}{\partial t}=\hat L S,
\label{eq3-4}
\end{equation}
and the probability density $\rho(q,p,t)=\psi^*(q,p,t)\psi(q,p,t)$ also
obeys the Liouville equation (\ref{eq3-1}). Therefore we can introduce 
a Hilbert space of square integrable complex functions $\psi(q,p,t)$, 
equip it with the the inner product
\begin{equation}
<\psi|\phi>=\int dqdp\, \psi^*(p,q,t)\phi(p,q,t),
\label{eq3-5}
\end{equation}
and then we recover the rules that are usually associated with quantum 
mechanics. Namely, observables are represented by Hermitian operators and the 
expectation value of an observable $\hat \Lambda$ is given by
\begin{equation}
\bar \Lambda(t)=\int dqdp\,\psi^*(q,p,t) \hat \Lambda \psi(q,p,t).
\label{eq3-6}
\end{equation}
If $\varphi_\lambda(q,p,t)$ is an eigenstate of the observable $\hat \Lambda$,
$\hat \Lambda \varphi_\lambda(q,p,t)=\lambda\,\varphi_\lambda(q,p,t)$, then
the probability $P(\lambda)$ that the outcome of a measurement of 
$\hat \Lambda$ on a classical mechanical system with the KvN wave function 
$\phi(p,q,t)$  results in the eigenvalue $\lambda$ is given by the usual Born 
rule
\begin{equation}
P(\lambda)=\int dqdp\,|\varphi^*_\lambda(q,p,t)\psi(q,p,t)|^2.
\label{eq3-7}
\end{equation}
There are two main differences from quantum mechanics. Firstly, and most 
importantly in the classical theory the operators for position and momentum 
do commute
\begin{equation}
[\hat q,\,\hat p]=0.
\label{eq3-8}
\end{equation}
In the Hilbert space formalism outlined above, these operators are realized 
as multiplicative operators
\begin{equation}
\hat q \,\psi(q,p,t)=q\,\psi(q,p,t),\;\;\; \hat p\, \psi(q,p,t)=p\,\psi(q,p,t).
\label{eq3-9}
\end{equation}
The second important difference is that the ``Hamiltonian'' (Liouville 
operator) (\ref{eq3-2}) that defines the time evolution of the KvN wave 
function is linear in spatial derivatives. This is quite unusual in quantum 
mechanics and such type of dynamical evolution was attributed to quantum 
systems that allow a genuine quantum chaos to emerge \cite{3-11}.

Thanks to the imaginary unit $i$ (and that's the reason why it was 
introduced), the Liouville operator $\hat L$ is Hermitian, and thus generates 
a unitary evolution, with respect to the inner product (\ref{eq3-5}):
\begin{equation}
\int dpdq\,\varphi^*(p,q,t) \hat L \psi(p,q,t)=
\int dpdq\,(\hat L \varphi)^*(p,q,t)\psi(p,q,t).
\label{eq3-10}
\end{equation}
This can be proved through an integration by parts under reasonable 
assumptions about the Hamiltonian, namely the equality of mixed derivatives
$\frac{\partial^2 H}{\partial q \partial p}=\frac{\partial^2 H}{\partial p 
\partial q}$. At that we assume that the wave functions $\varphi(q,p,t)$ and
$\psi(q,p,t)$, being square integrable, vanish sufficiently fast at $q,p\to
\pm\infty$. 

There are some mathematical subtleties here, however. Strictly speaking, 
not every square integrable function vanishes at infinity. The example is 
\cite{3-12A,3-12} $f(x)=x\exp{(-x^8\sin^2{x})}$ which is square integrable but 
even not bounded at infinity.\footnote{There is a typo in both \cite{3-12A} and
\cite{3-12}: $f(x)=x^2\exp{(-x^8\sin^2{x})}$ is not square integrable. In 
general  $f(x)=|x|^a\exp{(-|x|^b\sin^2{x})}$ is square integrable only if 
$2a-b/2<-1$. We thank Iosif Pinelis for clarifying this issue to us.}
According to the Hellinger-Toeplitz theorem
\cite{3-13}, everywhere defined Hermitian operator is necessarily bounded.
Position and momentum operators are clearly unbounded. So is the Liouville 
operator. Therefore, the rigorous mathematical formulation of classical
mechanics in the Hilbert space KvN formalism is not as a simple task as 
naively can appear. However, these mathematical subtleties and difficulties
are not characteristic of only KvN mechanics and is already present in 
ordinary quantum mechanics \cite{3-12}. The formalism of rigged Hilbert 
spaces  \cite{3-6} can provide a possible, although a rather sophisticated 
solution.

What the Hilbert space KvN formalism corresponds to the usual classical
mechanics is most easily seen in the Heisenberg picture of time evolution.
In the Schr\"{o}dinger picture of evolution assumed above the operators
are time-independent while the wave function evolves unitarily according to
\begin{equation}
\psi(q,p,t)=e^{-i\hat L t}\psi(q,p,0).
\label{eq3-11}
\end{equation}
On the contrary, in the Heisenberg picture wave functions are assumed to be
time-independent and all time dependencies of mean values  of physical 
quantities are incorporated in the time evolution of operators according to
\begin{equation}
\hat \Lambda(t)=e^{i\hat L t}\hat \Lambda(0)e^{-i\hat L t}.
\label{eq3-12}
\end{equation}
Equation of motion that follows from (\ref{eq3-12}) is
\begin{equation}
\frac{d\hat \Lambda(t)}{dt}=i[\hat L,\hat \Lambda(t)].
\label{eq3-13}
\end{equation}
Namely, for multiplicative position and momentum operators we get
\begin{equation}
\frac{d q}{dt}=i[\hat L,q]=\frac{\partial H(q,p,t)}{\partial p},\;\;\;
\frac{d p}{dt}=i[\hat L,p]=-\frac{\partial H(q,p,t)}{\partial q},
\label{eq3-14}
\end{equation}
which are nothing but the Hamilton's equations.

Alternatively, to show that KvN formalism corresponds to the usual Newtonian
mechanics, we can apply method of characteristics in the Schr\"{o}dinger 
picture \cite{3-13A}. Let us consider a curve $(q(\alpha),p(\alpha),
t(\alpha))$ in the extended phase space, parametrized by a real parameter
$\alpha$. Along this curve 
\begin{equation}
\frac{d\psi}{d\alpha}=\frac{\partial\psi}{\partial t}\frac{dt}{d\alpha}+
\frac{\partial\psi}{\partial q}\frac{dq}{d\alpha}+
\frac{\partial\psi}{\partial p}\frac{dp}{d\alpha},
\label{eq3-add1}
\end{equation}
and if the curve is chosen in such a way  that
\begin{equation}
\frac{dt}{d\alpha}=1,\;\;\;\frac{dq}{d\alpha}=\frac{\partial H}{\partial p},
\;\;\; \frac{dp}{d\alpha}=-\frac{\partial H}{\partial q},
\label{eq3-add2}
\end{equation}
we will get 
\begin{equation}
\frac{d\psi}{d\alpha}=-i\left (i\frac{\partial\psi}{\partial t}-\hat{L}\psi
\right )=0,
\label{eq3-add3}
\end{equation}
according to the Liouville equation (\ref{eq3-3}).

As we see from (\ref{eq3-add2}), the parameter $\alpha$ essentially coincides
with time and the characteristics of the Liouville equation (\ref{eq3-3}) are
just classical Newtonian trajectories in the extended phase space. Moreover,
the KvN wave function $\psi(q,p,t)$ remains constant along these 
trajectories. Thus delta-function initial date, with definite initial values 
of $(q_0,p_0,t_0)$, will be transported along Newtonian trajectories
$(q(t),p(t),t)$, as expected for a classical point particle.

In fact the Liouville operator (\ref{eq2-2}) is not uniquely defined in the 
KvN mechanics \cite{3-13B}. In particular, as it is clear from (\ref{eq3-14}),
we can add to the Liouville operator (\ref{eq3-2}) any function $F(q,p,t)$:
\begin{equation}
\hat{L}^\prime=\hat{L}+F(q,p,t),
\label{eq3-add4}
\end{equation}
without changing the Hamilton's equations (\ref{eq3-14}).

Of course, this gauge freedom in the choice of the Liouville operator is 
related to the invariance of the KvN probability density function under
the phase transformations
\begin{equation}
\psi^\prime(q,p,t)=e^{ig(q,p,t)}\psi(q,p,t).
\label{eq3-add5}
\end{equation}
Indeed, the new wave function $\psi^\prime(q,p,t)$ obeys the new Liouville 
equation
\begin{equation}
i\frac{\partial{\psi^\prime}}{\partial t}=\hat{L}^\prime\psi^\prime,
\label{eq3-add6}
\end{equation}
where
\begin{equation}
\hat{L}^\prime=e^{ig}\hat{L}e^{-ig}-\frac{\partial g}{\partial t}=
\hat{L}-\frac{\partial g}{\partial t}+\{H,g\}\equiv\hat{L}+F(q,p,t).
\label{eq3-add7}
\end{equation}
If evolution of the KvN wave function is determined by $\hat{L}^\prime$,
then along the Newtonian trajectories we will have
\begin{equation}
F(q,p,t)=-\left(\frac{\partial g}{\partial t}+\{g,H\}\right )=-
\frac{dg(q,p,t)}{dt},
\label{eq3-add8}
\end{equation}
and
\begin{equation}
\frac{d\psi^\prime}{dt}=-i\left (i\frac{\partial\psi^\prime}{\partial t}-
\hat{L}\psi^\prime\right )=-iF\psi^\prime=i\frac{dg}{dt}\psi^\prime,
\label{eq3-add9}
\end{equation}
which implies
\begin{equation}
\psi^\prime(q,p,t)=e^{i[g(q,p,t)-g(q_0,p_0,t_0)]}\psi^\prime(q_0,p_0,t_0).
\label{eq3-add10}
\end{equation}
That is, the KvN wave function no longer remains constant along Newtonian 
trajectories (along characteristics of the new  Liouville equation
(\ref{eq3-add6})), but the change affects only the phase of the wave function,
and such a change is irrelevant in the context of classical mechanics. 

An obvious difference between the KvN wave function and the true quantum wave 
function is the number of independent variables: KvN wave function depends
typically on the phase space variables $q,p$ and time, while quantum wave 
function typically depends on the configuration space variables ($q$ in our
case) and time. For an interesting perspective on the importance of this 
difference, see \cite{3-13B}. 

There were attempts to develop operator formulation of classical mechanics 
based on wave functions defined over configuration space, not over phase 
space \cite{3-13C,3-13D}. In such attempts Hamilton-Jacobi equation, not the 
Liouville equation, is used as a starting point. Although interesting, we will 
not pursue such approach in the present work.

A very interesting perspective on the KvN mechanics was given by Sudarshan
\cite{3-14,3-15}. Let us consider a quantum mechanical system with twice
as many degrees of freedom as our initial classical mechanical system. 
Namely, besides $\hat q$ and $\hat p$ operators, let us introduce new 
operators $\hat Q$ and $\hat P$ so that $(\hat q,\hat P)$ and $(\hat Q,
\hat p)$ form canonical pairs from the quantum mechanical point of view:
\begin{equation}
[\hat q,\hat P]=i\hbar,\;\;\;[\hat Q,\hat p]=i\hbar.
\label{eq3-15}
\end{equation}
Then in the $(q,p)$-representation, where $\hat q$ and $\hat p$ operators
are diagonal multiplicative operators, we will have
\begin{equation}
\hat P=-i\hbar\,\frac{\partial}{\partial q},\;\;\;{\mathrm and}\;\;\;
\hat Q=i\hbar\,\frac{\partial}{\partial p},
\label{eq3-16}
\end{equation}
so that the Liouville operator (\ref{eq3-2}) takes the form
\begin{equation}
\hat L=\frac{1}{\hbar}\left (\frac{\partial H}{ \partial q}\,\hat Q+
\frac{\partial H}{ \partial p}\,\hat P\right ),
\label{eq3-17}
\end{equation}
and the Liouville equation (\ref{eq3-3}) can be rewritten as a Schr\"{o}dinger
equation
\begin{equation} 
i\hbar \frac{\partial \psi}{\partial t}=\hat {\cal H} \psi,
\label{eq3-18}
\end{equation}
with the quantum Hamiltonian
\begin{equation} 
{\cal H}=\frac{\partial H}{ \partial q}\,\hat Q+
\frac{\partial H}{ \partial p}\,\hat P.
\label{eq3-19}
\end{equation}
The search for a hidden variable theory for quantum mechanics is a still 
ongoing saga \cite{3-16}. Here, thanks to Sudarshan (for earlier thoughts in 
this direction see \cite{3-17}) we have an amusing situation: classical 
mechanics, on the contrary, is interpreted as a hidden variable quantum 
theory! ``If we assume that not all quantum dynamical variables are actually 
observable, and if we set rules for distinguishing measurable from 
nonmeasurable operators, it is then possible to define a classical system as 
a special type of quantum system for which all measurable operators commute
`` \cite{3-18}.

What remains is to explain how Schr\"{o}dinger cat states is avoided in 
KvN mechanics: the superposition principle is the basic tenet of the 
quantum mechanics while in the classical realm the cat is either alive or 
dead, any superposition of these classical states does not make sense.

Of course, the fact that the amplitude and phase evolve independently,
equations (\ref{eq3-4}), already implies the absence of any interference
effects in KvN mechanics. However, this separation of the amplitude and phase
is an artifact of the $(q,p)$-representation. We can choose to work, for 
example, in the $(q,Q)$-representation instead \cite{3-10,3-19}. In this 
representation $\hat q$ and $\hat Q$ are simultaneously diagonal 
multiplicative operators, while $\hat p$ and $\hat P$ are differential 
operators:
\begin{eqnarray} &&
\hat q\,\psi(q,Q,t)=q\,\psi(q,Q,t),\hspace*{11mm}
\hat Q\,\psi(q,Q,t)=Q\,\psi(q,Q,t),
\nonumber \\ &&
\hat p\,\psi(q,Q,t)=-i\hbar \frac{\partial}{\partial Q}\,\psi(q,Q,t),\;\;
\hat P\,\psi(q,Q,t)=-i\hbar \frac{\partial}{\partial q}\,\psi(q,Q,t).
\label{eq3-20}
\end{eqnarray}
Wave functions in two representations are related by Fourier transform
(the same symbol $\psi$ is used for both the function and its Fourier 
transform for notational simplicity):
\begin{equation}
\psi(q,Q,t)=\frac{1}{\sqrt{2\pi}}\int dp\, e^{ipQ/\hbar}\psi(q,p,t).
\label{eq3-21}
\end{equation}
This follows from the following \cite{3-10}. If $|q,Q>$ are the simultaneous
eigenstates of the $\hat q$ and $\hat Q$ operators, while $|q,p>$ --- 
simultaneous eigenstates of the $\hat q$ and $\hat p$ operators:
\begin{eqnarray} &&
\hat q\,|q,Q>=q\,|q,Q>,\;\;\hat Q\,|q,Q>=Q\,|q,Q>,\nonumber \\ &&
\hat q\,|q,p>=q\,|q,p>,\;\;\;\;\,\hat p\,|q,p>=p\,|q,p>,
\label{eq3-22}
\end{eqnarray}
then we will have
\begin{eqnarray} &&
q<q^\prime,p^\prime|q,Q>=<q^\prime,p^\prime|\hat q|q,Q>=
q^\prime<q^\prime,p^\prime|q,Q>,\nonumber \\ &&
Q<q^\prime,p^\prime|q,Q>=<q^\prime,p^\prime|\hat Q|q,Q>=
-i\hbar\frac{\partial}{\partial p^\prime}<q^\prime,p^\prime|q,Q>,
\label{eq3-23}
\end{eqnarray}
which, together with the normalization condition
\begin{equation}
<q^\prime,Q^\prime|q,Q>=\delta(q^\prime-q)\delta(Q^\prime-Q),
\label{eq3-24}
\end{equation}
imply that in the $(q,p)$-representation the $|q,Q>$ state is given by the 
wave function
\begin{equation}
<q^\prime,p^\prime|q,Q>=\frac{1}{\sqrt{2\pi}}\,\delta(q^\prime-q)e^{-ip^\prime
Q/\hbar}.
\label{eq3-25}
\end{equation}
Because, like $|q,p>$ states, $|q,Q>$ states also form a complete set of 
orthonormal eigenstates in the KvN Hilbert space, we have
\begin{equation}
\psi(q,Q,t)=<q,Q|\psi(t)>=\int dq^\prime dp\, 
<q,Q|q^\prime,p><q^\prime,p|\psi(t)>,
\label{eq3-26}
\end{equation}
which, in light of (\ref{eq3-25}), is equivalent to (\ref{eq3-21}).

In the $(q,Q)$-representation and for the classical Hamiltonian $H=\frac{p^2}
{2m}+V(q)$, the quantum Hamiltonian (\ref{eq3-19}) takes the form
\begin{equation}
{\cal H}=\frac{dV}{dq}\,Q-\frac{\hbar^2}{m}\,\frac{\partial^2}{\partial q
\partial Q}.
\label{eq3-27}
\end{equation}
Then it follows from the Schr\"{o}dinger equation (\ref{eq3-18}) that the 
amplitude $A(q,Q)$ and the phase $\Phi(q,Q)$ of the KvN wave function 
$\psi(q,Q)=A\,e^{i\Phi/\hbar}$ evolve according to the equations
\begin{eqnarray} &&
\frac{\partial A}{\partial t}+\frac{1}{m}\left(\frac{\partial A}{\partial q}
\frac{\partial \Phi}{\partial Q}+\frac{\partial A}{\partial Q}
\frac{\partial \Phi}{\partial q}+A\,\frac{\partial^2 \Phi}{\partial Q
\partial q}\right)=0,\nonumber \\ &&
\frac{\partial \Phi}{\partial t}+\frac{1}{m}\left(\frac{\partial \Phi}
{\partial q}\frac{\partial \Phi}{\partial Q}-\frac{\hbar^2}{A}
\frac{\partial^2 A}{\partial Q\partial q}\right)+\frac{dV}{dq}\,Q=0.
\label{eq3-28}
\end{eqnarray}
As we see, in this representation the phase and amplitude are coupled in the 
equations of motion and their time evolutions become intertwined much like the
ordinary quantum mechanics. This is hardly surprising because, after all, the
encompassing  underlying system is quantum.

According to Sudarshan \cite{3-14,3-15}, it is the superselection principle
\cite{3-20,3-21} which kills the interference effects in the KvN mechanics.
In classical mechanics, observables are functions of the phase space variables
$q$ and $p$. Therefore, $\hat q$ and $\hat p$ commute with all classical 
observables and thus trigger a superselection mechanism which render the 
relative phase between different superselection sectors unobservable. Indeed, 
let
\begin{equation}
|\psi>=\alpha|p,q>+\beta|p^\prime,q^\prime>,
\label{eq3-29}
\end{equation}
with $|\alpha|^2+|\beta|^2=1$, be a seemingly coherent superposition of 
different eigenstates of $\hat q$ and $\hat p$. As we assume that $|p,q>$
form a complete set of orthonormal states and an observable $\hat \Lambda$
commutes with $\hat q$ and $\hat p$, $|p,q>$ is an eigenstate of 
$\hat \Lambda$ also and thus $<p^\prime,q^\prime|\hat \Lambda|p,q>=0$.
Therefore, for the mean value of the observable $\hat \Lambda$ in the state
$|\psi>$ we get
\begin{equation}
\bar \Lambda=<\psi|\hat \Lambda|\psi>=|\alpha|^2 <p,q|\hat \Lambda|p,q>+
|\beta|^2 <p^\prime,q^\prime|\hat \Lambda|p^\prime,q^\prime>.
\label{eq3-30}
\end{equation}
As we see, all interference effects are gone and the mean value is the same 
as if we had an incoherent mixture of the states $|p,q>$ and 
$|p^\prime,q^\prime>$ described by the diagonal density matrix
\begin{equation}
\hat \rho=|\alpha|^2|\,|p,q><p,q|+|\beta|^2|\,|p^\prime,q^\prime>
<p^\prime,q^\prime|.
\label{eq3-31}
\end{equation}
However, this use of the superselection principle in KvN mechanics differs
from its conventional use in one essential aspect \cite{3-10,3-14}. In quantum 
mechanics time evolution is governed by Hamiltonian which is by itself
an observable. As a result all time evolution takes place in one 
superselection sector and we have genuine superselection rules that the 
eigenvalues of the superselecting operators cannot be changed during the time 
evolution. Of course, in the case of KvN mechanics this would be 
a catastrophe because it would imply that $q$ and $p$ cannot change during
the time evolution. Fortunately, the quantum Hamiltonian (\ref{eq3-19}) is
not a classical observable, because it contains unobservable hidden quantum
variables $\hat Q$ and $\hat P$. As a result (\ref{eq3-19}) does not commute 
with $\hat q$ and $\hat p$ operators and thus can generate a transition from 
one eigenspace of these superselection operators to the other.

In conclusion of our mini review of the KvN mechanics, let's indicate some
further references \cite{3-22,3-23,3-24,3-25,3-26,3-27,3-28,3-29,3-30} where
an interested reader can find modern developments related to the KvN 
mechanics. Although it is not directly related to the KvN mechanics, let us 
also mention that in 1973  Martin, Siggia and  Rose proposed  an operator 
formalism  for certain types of classical systems that is very similar to 
the Schwinger formulation of quantum field theory \cite{3-31}.

\section{Modification of classical mechanics}
Modified commutation relations alone are not enough to derive physical 
meaning. Many attempts were made  to define the dynamics of quantum systems
and their observables in the presence of a minimal length, but this research 
field is still far from being as logically consistent and mature as the 
ordinary quantum mechanics is \cite{2-37}. In any case, the minimal length 
modification of quantum mechanics entails the corresponding modification of 
classical mechanics, as the former is considered as the $\hbar \to 0$ limit 
of the latter (see, however, \cite{4-1}). 

Usually the modification of classical mechanics is obtained from the 
corresponding modification of quantum mechanics  by replacing modified 
commutators with modified Poisson brackets \cite{2-35,4-2}:
\begin{equation}
\frac{1}{i\hbar}\,[\hat x_i,\,\hat p_j] \to \{x_i,\,p_j\}.
\label{eq4-1}
\end{equation}
At that the modified Poisson bracket  of arbitrary functions $F$ and $G$ of 
the coordinates and momenta are defined as \cite{2-35}
\begin{equation}
\{F,\,G\}=\left (\frac{\partial F}{\partial x_i}\,\frac{\partial G}
{\partial p_j}-\frac{\partial F}{\partial p_i}\,\frac{\partial G}
{\partial x_i}\right )\{x_i,\,p_j\}+\frac{\partial F}{\partial x_i}\,
\frac{\partial G}{\partial x_j}\,\{x_i,\,x_j\}.
\label{eq4-2}
\end{equation}
Correspondingly, the classical equations of motion have the form
\begin{equation}
\dot x_i=\{x_i,\,H\}=\{x_i,\,p_j\}\,\frac{\partial H}{\partial p_j}+
\,\{x_i,\,x_j\}\,\frac{\partial H}{\partial x_j},\;\;
\dot p_i=\{p_i,\,H\}=-\{x_i,\,p_j\}\,\frac{\partial H}{\partial x_j}.
\label{eq4-3}
\end{equation}
A number of classical mechanics problem was studied within this scenario
\cite{4-3,4-4,4-5,4-6,4-7,4-8,4-9,4-10}. Koopman-von Neumann mechanics,
however, provides a different and in our opinion more interesting perspective
on the Planck scale deformation of classical mechanics. 

The main idea is the following: modification of the commutation relations,
for example in the form of (\ref{eq2-18}), in the encompassing (in the 
Sudarshan sense) quantum system will alter classical dynamics in the 
$(q,p)$ classical subspace. 

For simplicity, let us consider a one-dimensional classical harmonic
oscillator with the Hamiltonian
\begin{equation}
H=\frac{1}{2m}(p^2+m^2\omega^2\,q^2).
\label{eq4-4}
\end{equation}
In the Sudarshan-encompassing two-dimensional quantum system we can 
identify $x_1=q,\,x_2=Q,\,p_1=P,\,p_2=p$. Then Snyder commutation relations
(\ref{eq2-17}) take the form
\begin{eqnarray} &&
[\hat q,\,\hat P]=i\hbar (1+\beta^\prime \hat P^2),\;\;
[\hat q,\,\hat p]=[\hat Q,\,\hat P]=i\hbar\beta^\prime\hat p\hat P,\;\;
\nonumber \\ &&
[\hat Q,\,\hat p]=i\hbar (1+\beta^\prime \hat p^2),\;\;
[\hat q,\,\hat Q]=i\hbar\beta^\prime(\hat q\hat p-\hat Q\hat P),\;\;
[\hat p,\,\hat P]=0.
\label{eq4-5}
\end{eqnarray}
The first surprise is that $\hat q$ and $\hat p$ cease to be commuting.
According to (\ref{eq2-11}), the corresponding uncertainty relation is
\begin{equation}
\Delta q\Delta p\ge\frac{\hbar\beta^\prime}{2}\left (\Delta(p,P)+<p><P>
\right).
\label{eq4-6}
\end{equation}
As we see the sharply defined classical trajectories cease to exist in 
the $(q,p)$ phase space, much like the quantum case. The constant 
$\hbar\beta^\prime$, that governess the fuzziness of the ``classical''
$(q,p)$ phase space, is induced, we believe, by the quantum gravity/string 
theory effects at the Planck scale. Then
\begin{equation}
\hbar\beta^\prime\sim \frac{l_P^2}{\hbar}=\frac{G}{c^3},
\label{eq4-7}
\end{equation}
and we see that it is expected not to depend on $\hbar$! Classical trajectories
will be lost even in the hypothetical world with $\hbar=0$, provided the
Newton constant $G$ is not zero and the universal velocity $c$ is not 
infinity. However, Our troubles with classicality don't end here. The quantum
Hamiltonian (\ref{eq3-19}) that corresponds to (\ref{eq4-4}) has the form
\begin{equation}
{\cal H}=\frac{1}{m}\left(\hat p\hat P+m^2\omega^2 \hat q \hat Q\right ),
\label{eq4-8}
\end{equation}
indicating, according to (\ref{eq4-5}), the following equations of motion
(in the Heisenberg picture) for ``classical'' variables $\hat q$ and
$\hat p$ :
\begin{eqnarray} &&
\frac{d\hat q}{dt}=\frac{i}{\hbar}[{\cal H},\hat q]=\frac{\hat p}{m}\left (
1+2\beta^\prime\hat P^2\right )+
\beta^\prime m\omega^2\hat q(\hat q\hat p-\hat Q\hat P),\;\;\;
\nonumber \\ &&
\frac{d\hat p}{dt}=\frac{i}{\hbar}[{\cal H},\hat p]=-m\omega^2\hat q
\left (1+\beta^\prime \hat p^2\right )-
\beta^\prime m\omega^2\hat p\hat P\hat Q.
\label{eq4-9}
\end{eqnarray}
As one would expect from the beginning, the equations are modified. What was 
probably unexpected is that the additional terms depend on the hidden
variables $\hat Q$ and $\hat P$! 

Analogous equations hold for ``hidden'' variables $\hat Q$ and
$\hat P$ :
\begin{eqnarray} &&
\frac{d\hat Q}{dt}=\frac{i}{\hbar}[{\cal H},\hat Q]=\left (
1+2\beta^\prime\hat p^2\right )\frac{\hat P}{m}-
\beta^\prime m\omega^2(\hat q\hat p-\hat Q\hat P)\hat Q,\;\;\;
\nonumber \\ &&
\frac{d\hat P}{dt}=\frac{i}{\hbar}[{\cal H},\hat P]=-m\omega^2
\left (1+\beta^\prime \hat P^2\right )\hat Q-
\beta^\prime m\omega^2\hat q\hat p\hat P.
\label{eq4-10}
\end{eqnarray}
If the modification considered emerges from the Planck scale effects, the 
natural scale for new phenomenological constants, like  $\beta^\prime$, is
$\beta^\prime\sim l_P^2/\hbar^2=1/p_P^2$, where $p_P$ is the Planck momentum.
Therefore the correction terms in (\ref{eq4-5}) and  (\ref{eq4-9}) are
significant only when the momenta involved are of the order of Planck momentum.
Let us suppose that this is indeed so for $(q,p)$ classical sector related 
momenta, while the hidden $(Q,P)$ sector related momenta for some reason 
remains much smaller, so that we can discard hidden variables $(Q,P)$ in
(\ref{eq4-5}) and  (\ref{eq4-9}). Then we still regain the classical sector
($\hat q$ and $\hat p$ will commute in this approximation), but the classical 
equations of motion will be modified (as $\hat q$ and $\hat p$ do commute, we 
write equations of motion for $q$ and $p$ considering them as real numbers, 
not operators):
\begin{equation}
\frac{dq}{dt}=\left (1+\beta^\prime m^2\omega^2 q^2\right )\frac{p}{m},\;\;\;
\frac{dp}{dt}=-m\omega^2 \left (1+\beta^\prime p^2\right ) q.
\label{eq4-11}
\end{equation}
Equations (\ref{eq4-11}) are non-linear oscillator equations  of the type 
introduced in \cite{4-11,4-12} that  model generalized one-dimensional 
harmonic oscillators in several important dynamical systems:
\begin{equation}
\frac{dq}{dt}=f(q)\,p,\;\;\;\frac{dp}{dt}=-g(p)\,q,
\label{eq4-12}
\end{equation}
where $f(q)$ and $g(p)$ functions, with the conditions $f(0)>0,\,g(0)>0$, are 
assumed to be continuous with continuous first derivatives.

On the other hand, (\ref{eq4-11}) can be rewritten as an second order
differential equation and the result is
\begin{equation}
\ddot q -\frac{\beta^\prime m^2\omega^2 q}{1+\beta^\prime m^2\omega^2 q^2}
\,\dot q^2+\omega^2(1+\beta^\prime m^2\omega^2 q^2)q=0.
\label{eq4-13}
\end{equation}
This equation is of the type of quadratic Li\'{e}nard equation. The general
quadra\-tic Li\'{e}nard equation, used in a vast range of applications, has 
the form
\begin{equation}
\ddot q+f(q)\,\dot q^2+g(q)=0,
\label{eq4-14}
\end{equation}
where $f(q)$ and $g(q)$ are arbitrary functions that do not vanish 
simultaneously \cite{4-13,4-13A}.

The equation (\ref{eq4-13}) (and in general the system (\ref{eq4-12})
\cite{4-11}) admits a first integral which can be found as follows. From
(\ref{eq4-11}) we have
\begin{equation}
\frac{dp}{dq}=-\frac{m^2\omega^2(1+\beta^\prime p^2)q}
{(1+\beta^\prime m^2\omega^2 q^2)p}.
\label{eq4-15}
\end{equation}
The variables $q$ and $p$ in this differential equation can be separated and 
we get after the integration
$$\ln{(1+\beta^\prime p^2)}+\ln{(1+\beta^\prime m^2\omega^2 q^2)}=
\mathrm{constant},$$
which implies that
\begin{equation}
(1+\beta^\prime p^2)(1+\beta^\prime m^2\omega^2 q^2)=1+\beta^\prime A,
\label{eq4-16}
\end{equation}
where $A$ is some constant. For definiteness, let us assume the following 
initial values
\begin{equation}
q(0)=0,\;\;\;p(0)=p_0>0.
\label{eq4-17}
\end{equation}
Then $A=p_0^2$.

Another interesting property of the equation (\ref{eq4-13}) is that it 
corresponds to a Lagrangian system with a position dependent mass (in fact, 
any quadratic Li\'{e}nard equation has such a property \cite{4-14}). Indeed,
the Lagrangian 
\begin{equation}
{\cal L}=\frac{1}{2}\mu(q)\,\dot q^2-V(q)
\label{eq4-18}
\end{equation}
leads to the Euler-Lagrange equation
$$\ddot q+\frac{\mu^\prime}{2\mu}\,\dot q^2+\frac{V^\prime}{\mu}=0.$$
Here the prime denotes differentiation with respect to $q$. Comparing with
(\ref{eq4-13}), we get the identifications
\begin{equation}
\frac{\mu^\prime}{2\mu}=-\frac{\beta^\prime m^2\omega^2 q}
{1+\beta^\prime m^2\omega^2 q^2},\;\;\;
V^\prime=\mu\,\omega^2 q\left (1+\beta^\prime m^2\omega^2 q^2\right).
\label{eq4-19}
\end{equation}
These differential equations can be integrated and we get
\begin{equation}
\mu(q)=\frac{m}{1+\beta^\prime m^2\omega^2 q^2},\;\;\;
V(q)=\frac{1}{2}m\omega^2 q^2.
\label{eq4-20}
\end{equation}
Comparing the conserved energy $E=\frac{p^2}{2\mu}+V$ with (\ref{eq4-16}), we 
see that the first integral (\ref{eq4-16}) represents just the energy 
conservation and $A=2mE$.

Thanks to the first integral (\ref{eq4-16}), the equation of motion 
(\ref{eq4-13}) of the Planck scale deformed harmonic oscillator can be solved 
in a quadrature. Namely, from (\ref{eq4-16}), assuming (\ref{eq4-17}) initial
conditions, we get
$$p=\sqrt{\frac{p_0^2-m^2\omega^2 q^2}{1+\beta^\prime m^2\omega^2 q^2}}.$$
On the other hand, it follows from the first equation of (\ref{eq4-11}) that
$$p=\frac{m\dot q}{1+\beta^\prime m^2\omega^2 q^2}.$$
Combining these two expressions of $p$, we get
\begin{equation}
m\frac{dq}{dt}=\sqrt{(p_0^2-m^2\omega^2 q^2)(1+\beta^\prime m^2\omega^2 q^2)}.
\label{eq4-21}
\end{equation}
Introducing a new variable $z=m\omega q/p_0$, we can integrate (\ref{eq4-21})
as follows:
\begin{equation}
\omega t=\int\limits_0^{m\omega q/p_0}\frac{dz}{\sqrt{(1-z^2)(1+
\beta^\prime p_0^2 z^2)}}=\int\limits_0^u\frac{d\theta}{\sqrt{1+
\beta^\prime p_0^2\sin^2{\theta}}},
\label{eq4-22}
\end{equation} 
where $\sin{u}=m\omega q/p_0$ and at the last step we have made another
change of the integration variable, namely $z=\sin{\theta}$. The above integral
is the incomplete elliptic integral of the first kind. Its amplitude $u$
satisfies the equation $\sin{u}=\sn{(\omega t,i\sqrt{\beta^\prime}p_0)}$, where
$\sn{(\omega t,i\sqrt{\beta^\prime}p_0)}$ is the Jacobi sine elliptic function
with the imaginary modulus (assuming $\beta^\prime>0$). Therefore
\begin{equation}
q(t)=\frac{p_0}{m\omega}\,\sn{(\omega t,i\sqrt{\beta^\prime}p_0)}=
\frac{p_0}{m\omega\sqrt{1+\beta^\prime p_0^2}}\sd{\left(\sqrt{1+
\beta^\prime p_0^2}\,\omega t,\,\sqrt{\frac{\beta^\prime p_0^2}{1+
\beta^\prime p_0^2}}\right )},
\label{eq4-23}
\end{equation}
where at the last step we have used imaginary modulus transformation 
\cite{4-15}. Period of oscillations $T$, according to (\ref{eq4-22}), is given
by the relation
\begin{equation}
\frac{\omega T}{4}=\int\limits_0^{\pi/2}\frac{d\theta}{\sqrt{1+
\beta^\prime p_0^2\sin{\theta}^2}}=\K(i\sqrt{\beta^\prime}),
\label{eq4-24}
\end{equation}
where $\K$ is the complete elliptic integral of the first kind. Using again 
the imaginary modulus transformation, we get
\begin{equation}
T=\frac{4}{\omega\sqrt{1+\beta^\prime p_0^2}}\,\K{\left(\sqrt{\frac
{\beta^\prime p_0^2}{1+\beta^\prime p_0^2}}\right )}\approx \frac{2\pi}{\omega}
\left (1-\frac{\beta^\prime p_0^2}{4}\right ).
\label{eq4-25}
\end{equation} 
This reduction of the period of oscillations is similar to what was found in
\cite{4-9} within the framework of the one-dimensional Kempf modification of 
the commutation relations (a one-dimensional version of (\ref{eq2-18})) with 
the standard recipe of replacing commutators by Poisson brackets when 
considering a classical limit.

Let us also consider, as a second example, Kempf et al. modification
of the commutation relations (\ref{eq2-18}) with $\beta^\prime=2\beta$, so 
that the spatial geometry remains approximately commutative (at the first 
order in $\beta$). Then we will have (for our two-dimensional quantum system
the square of the momentum vector is $p^2+P^2$)
\begin{eqnarray} && 
[\hat q,\hat P]=i\hbar[1+\beta(\hat p^2+3\hat P^2)],\;\; [\hat q,\,\hat p]=
[\hat Q,\,\hat P]=i\hbar\,2\beta \hat p\hat P,\;\;
\nonumber \\ &&
[\hat Q,\,\hat p]=i\hbar[1+\beta(3\hat p^2+\hat P^2)],\;\;  
[\hat q,\,\hat Q]=0,\;\; [\hat p,\,\hat P]=0.
\label{eq4-26}
\end{eqnarray} 
Again $q$ and $p$ cease to be commuting. Then corresponding equations of motion
are
\begin{eqnarray} &&
\frac{d\hat q}{dt}=\frac{\hat p}{m}\left [1+\beta\left(\hat p^2+
5\hat P^2\right )\right],\;\;\;\; 
\nonumber \\ &&
\frac{d\hat p}{dt}=-m\omega^2 \left \{\hat q
\left[1+\beta(3\hat p^2+\hat P^2)\right ]+2\beta \,\hat p\,\hat P \hat Q
\right \},
\label{eq4-27}
\end{eqnarray}
and
\begin{eqnarray} &&
\frac{d \hat Q}{dt}=\left [1+\beta\left (5\hat p^2+\hat P^2\right )\right ]
\frac{\hat P}{m},\;\;\;\;
\nonumber \\ &&
\frac{d \hat P}{dt}=-m\omega^2\left\{\left[1+\beta
\left (\hat p^2+3\hat P^2\right )\right ]\hat Q+2\beta\,\hat q\hat p\hat P
\right \}.
\label{eq4-28}
\end{eqnarray}
As in the previous example, hidden variables $P$ and $Q$ do appear in the 
equations of motions of the ``classical'' sector due to the Plank scale 
modification of the commutation relations.

Let us emphasize that equations (\ref{eq4-27}) and (\ref{eq4-28}) (as well as
equations (\ref{eq4-9}) and (\ref{eq4-10}) earlier) describe a (deformed) 
quantum system (in the Heisenberg picture). The Sudarshan perspective on KvN 
mechanics is that quantum dynamics of a specific kind induces a classical 
dynamics in its subsystem of twice smaller dimension. Remaining variables
of the quantum system which are outside of this classical subsystem remain 
hidden in the sense that they don't influence classical dynamics. Our results
indicate that this is no longer true if gravity is taken into account, because 
gravity, due to its universal character, couples to hidden variables too and 
as a result these variables begin to couple to the "classical" variables.

Strictly speaking, in such a situation it is not clear if one can speak 
about deformed classical mechanics at all because the setup now is inherently 
quantum mechanical. In the next section we analyze the corresponding (deformed)
quantum system in the Schrodinger picture to find some corrections to the 
classical evolution. Such an approach requires some assumption about the 
initial wave function and cannot be too general without further specifying the 
nature of the quantum system under consideration and of the corresponding 
``hidden'' variables.

However, in our opinion, one can still speak about deformed classical mechanics
in situations when the ``hidden'' variables in some sense remain small. 
Equations (\ref{eq4-10}) and (\ref{eq4-28}) show that if ``hidden'' variables
are initially zero then they will remain to be zero all the time. Of course 
more physical situation is that ``hidden'' variables are initially not 
strictly zero but small. In this case they don't remain small all the time 
because of the resonance coupling to the ``classical'' variables. However this 
coupling is controlled by a small parameter $\beta p^2$ and the corresponding 
rising time can be estimated as being $(\beta p^2 \omega)^{-1}$, and for all 
classical mechanical harmonic oscillator systems, considered in the context of 
potentially experimentally detectable deviations from classical mechanics due 
to quantum gravity effects, this rising time is several orders of magnitude 
larger than the age of the universe \cite{6-4}.
 
In situations when the effects of the hidden variables $P$ and $Q$ can be 
approximately discarded, the classical equations of motion became
\begin{equation}
\dot q=\left (1+\beta p^2\right )\frac{p}{m},\;\;\;
\dot p=-m\omega^2\left (1+3\beta p^2\right )q.
\label{eq4-29}
\end{equation}
This system is no longer of the type (\ref{eq4-12}). Nevertheless, it gives
a second order differential equation for the variable $p$ which is of the
quadratic Li\'{e}nard type:
\begin{equation}
\ddot p-\frac{6\beta p}{1+3\beta p^2}\,\dot p^2+\omega^2\left (1+\beta p^2
\right )\left ( 1+3\beta p^2\right )p=0.
\label{eq4-30}
\end{equation}
Variable mass system (in the $p$-space) with the Lagrangian
$${\cal L}=\frac{1}{2}\mu(p)\dot p^2-V(p),$$
which is equivalent to (\ref{eq4-30}), is characterized by
\begin{equation}
\mu(p)=\frac{m}{(1+3\beta p^2)^2},\;\;\;V(p)=\frac{m\omega^2}{6}\left [
p^2+\frac{2}{3\beta}\,\ln{\left(1+3\beta p^2\right )}\right ].
\label{eq4-31}
\end{equation}
Of course the integration constant in the potential $V(p)$ is irrelevant and 
it was chosen in such a way that when $\beta=0$ the Lagrangian ${\cal L}$ 
becomes the ordinary harmonic oscillator Lagrangian. The corresponding 
conserved ``energy'' $\frac{1}{2}\mu(p) \dot p^2+V(p)$ gives a first integral
\begin{equation}
\frac{1}{2}\,\frac{m}{(1+3\beta p^2)^2}\,\dot p^2+\frac{m\omega^2}{6}\left [
p^2+\frac{2}{3\beta}\,\ln{\left(1+3\beta p^2\right )}\right ]=m^2\omega^2 E,
\label{eq4-32}
\end{equation}
where $E$ is some constant.

Period of oscillations that follows from (\ref{eq4-32}) is
\begin{eqnarray} &&
T=4\int\limits_0^{p_0}\frac{dp}{\sqrt{\frac{2}{\mu}(m^2 \omega ^2 E-V(p))}}=
\nonumber \\ &&
\frac{4}{\omega} \int\limits_0^{p_0}\frac{dp}{(1+3\beta p^2)\sqrt{2mE-
\frac{1}{3}\left[p^2+\frac{2}{3\beta}\ln{(1+3\beta p^2)}\right ]}}.
\label{eq4-33}
\end{eqnarray}
At the first order in $\beta$, and assuming $\dot p(0)=0,\,p(0)=p_0$ initial
conditions, we have $E=\frac{p_0^2}{2m}(1-\beta p_0^2)$ and
\begin{eqnarray} &&
T\approx\frac{4}{\omega} \int\limits_0^{p_0}\frac{dp}{(1+3\beta p^2)\sqrt{
(p_0^2-p^2)\left[1-\beta(p_0^2+p^2)\right ]}}\approx 
\nonumber \\ && \frac{4}{\omega} 
\int\limits_0^{p_0}\frac{dp}{\sqrt{p_0^2-p^2}}\left [1-\frac{\beta}{2}
(5p^2-p_0^2)\right ]=\frac{2\pi}{\omega}\left (1-\frac{3\beta  p_0^2}{4}
\right ).
\label{eq4-34}
\end{eqnarray}
At the final step, we have used elementary integrals
$$\int\limits_0^{p_0}\frac{dp}{\sqrt{p_0^2-p^2}}=\frac{\pi}{2},\;\;\;
\int\limits_0^{p_0}\sqrt{p_0^2-p^2}\,dp=\frac{\pi p_0^2}{4}.$$

\section{Observable effects of ``hidden'' variables}
The introduction of hidden variables $(Q,P)$ in the original Koopman-von 
Neumann context is a purely formal move which acquires some substantially
in the Sudarshan perspective, because now they are considered as degrees of
freedom of an encompassing quantum system. However due to very specific 
character of the assumed quantum evolution (the quantum Hamiltonian 
(\ref{eq3-19}) is linear in $Q$ and $P$ variables), the dynamics in the 
classical $(q,p)$ sector remains isolated from the influence of what happens
in the hidden $(Q,P)$ sector. As we have seen, this is no longer true when
the quantum system is deformed due to the expected Planck scale quantum 
gravity effects. Because of an universal character of gravity, hidden variables
$Q$ and $P$ cease to be completely hidden for classical observers as they 
appear in the evolution equations of the ``classical'' $q$ and $p$ variables. 
The natural question then is to what extent the observable effects of the 
hidden variables $Q$ and $P$ can be neglected in the classical regime.

Not much can be said about this question without specifying the concrete nature
of the hidden variables and of the corresponding encompassing quantum system.
However some understanding can be gained by an approximate solution of the
(deformed) quantum mechanical problem with Hamiltonian (\ref{eq4-8}) and 
deformed commutation relations (\ref{eq4-26}).

First of all let's make a canonical transformation
\begin{equation}
\hat q=\frac{1}{\sqrt{2}}(\hat q_1-\hat q_2),\;\;\;
\hat Q=\frac{1}{\sqrt{2}}(\hat q_1+\hat q_2),\;\;\;
\hat p=\frac{1}{\sqrt{2}}(\hat p_1+\hat p_2),\;\;\;
\hat P=\frac{1}{\sqrt{2}}(\hat p_1-\hat p_2).
\label{eq6-1}
\end{equation}
The transformation (\ref{eq6-1}) is canonical in the sense that the new 
operators
\begin{equation}
\hat q_1=\frac{1}{\sqrt{2}}(\hat q+\hat Q),\;\;\;
\hat q_2=\frac{1}{\sqrt{2}}(\hat Q-\hat q),\;\;\;
\hat p_1=\frac{1}{\sqrt{2}}(\hat p+\hat P),\;\;\;
\hat p_2=\frac{1}{\sqrt{2}}(\hat p-\hat P)
\label{eq6-2}
\end{equation}
obey the very same Kempf et al. commutation relations (to first order in 
$\beta$):
\begin{equation}
[\hat q_i,\,\hat q_j]=0,\;\;\;[\hat q_i,\,\hat p_j]=i\hbar\left [\delta_{ij}+
\beta (\hat p_1^2+\hat p_2^2)\delta_{ij}+2\beta\,\hat p_i\hat p_j\right ],
\;\;\; [\hat p_i,\,\hat p_j]=0.
\label{eq6-3}
\end{equation}
If we introduce an auxiliary ``low energy momentum'' operators \cite{6-1,6-2}
$\hat \pi_i$ which obey the canonical commutation relations $[\hat q_i,\,\hat 
\pi_j]=i\hbar\delta_{ij}$, $[\hat \pi_i,\,\hat \pi_j]=0$, then to the first 
order in $\beta$ the ``high energy momentum'' operators $\hat p_i$ can be 
expressed through them as follows  \cite{6-1,6-2}
\begin{equation}
\hat p_i=\hat \pi_i\left[1+\beta (\hat \pi_1^2+\hat \pi_2^2)\right].
\label{eq6-4}
\end{equation}
Because of (\ref{eq6-1}) and (\ref{eq6-4}), the quantum Hamiltonian 
(\ref{eq4-8}) takes the form
\begin{equation}
{\cal H}={\cal H}_1-{\cal H}_2,\;\;\;{\cal H}_i=\frac{1}{2m}\left[\hat \pi_i^2
+m^2\omega^2\hat q_i^2+2\beta \hat\pi_i^4\right ].
\label{eq6-5}
\end{equation} 
Correspondingly, we first consider a quantum mechanical problem of perturbed
harmonic oscillator with the Hamiltonian
\begin{equation}
\hat h=\frac{1}{2}(\hat\eta^2+\hat \xi^2)+\alpha\hat\eta^4=\hat h_0+
\alpha\hat\eta^4,
\label{eq6-6}
\end{equation}
where we have introduced dimensionless variables \cite{6-3}
\begin{eqnarray} &&
\hat\xi=\sqrt{\frac{m\omega}{\hbar}}\,\hat q,\;\;\;
\hat\eta=\sqrt{\frac{1}{\hbar m\omega}}\,\hat \pi,\;\;\;
\nonumber \\ &&
\hat h=\frac{1}{\hbar \omega}\left (\frac{\hat\pi^2}{2m}+\frac{\beta}{m}
\hat\pi^4+\frac{1}{2}m\omega^2\hat q^2\right ),\;\;\;
\alpha=\beta m\hbar\omega.
\label{eq6-7}
\end{eqnarray}
The wave functions and the corresponding eigenvalues of the unperturbed
harmonic oscillator problem are well known \cite{6-3}:
\begin{equation}
\psi_n^{(0)}(\xi)=\frac{1}{\sqrt{2^nn!\sqrt{\pi}}}\,e^{-\xi^2/2}
H_n(\xi),\;\;\;\epsilon_n^{(0)}=n+\frac{1}{2},
\label{eq6-8}
\end{equation}
$H_n(\xi)$ being the Hermite polynomials. Then the standard time-independent 
perturbation theory (see, for example, \cite{6-3}) can be used to get the
first order corrections to them. In particular, $\epsilon_n^{(1)}=<n^{(0)}|
\alpha\hat\eta^4|n^{(0)}>$. While calculating this quantity, we can use zeroth 
order Schr\"{o}dinger equation $\hat h_0|n^{(0)}>=\epsilon_n^{(0)}|n^{(0)}>$ 
to get $\hat\eta^2|n^{(0)}>\approx (2\epsilon_n^{(0)}-\hat\xi^2)|n^{(0)}>$ and 
correspondingly
\begin{eqnarray} &&
\epsilon_n^{(1)}\approx\alpha<n^{(0)}|(2\epsilon_n^{(0)}-\hat\xi^2)^2|n^{(0)}>=
\alpha\left[4\epsilon_n^{(0)\,2}-4\epsilon_n^{(0)}<n^{(0)}|\hat\xi^2|n^{(0)}>+
\right .\nonumber \\ &&
\left . <n^{(0)}|\hat\xi^4|n^{(0)}>\right ].
\label{eq6-9}
\end{eqnarray}
Matrix elements of $\hat\xi^k$, $k=2,4,\ldots$, can be calculated thanks to the
orthonormality property $<n^{(0)}|m^{(0)}>=\delta_{n,m}$ and the recurrence 
relation  \cite{6-3}
\begin{equation}
\hat\xi|n^{(0)}>=\sqrt{\frac{n+1}{2}}\,|(n+1)^{(0)}>+\;\sqrt{\frac{n}{2}}\,
|(n-1)^{(0)}>.
\label{eq6-10}
\end{equation}
As a result, we get
\begin{equation}
<n^{(0)}|\hat\xi^2)|n^{(0)}>=n+\frac{1}{2},\;\;\; 
<n^{(0)}|\hat\xi^4)|n^{(0)}>=\frac{1}{4}(6n^2+6n+3),
\label{eq6-11}
\end{equation} 
and
\begin{equation}
\epsilon_n^{(1)}=\frac{\alpha}{4}(6n^2+6n+3).
\label{eq6-12}
\end{equation}
The first-order correction to the state vector has the form
\begin{eqnarray} &&
|n^{(1)}>=\alpha\sum\limits_{m\ne n}\frac{<m^{(0)}|\hat\eta^4|n^{(0)}>}
{\epsilon_n^{(0)}-\epsilon_m^{(0)}}\,|m^{(0)}>=
\nonumber \\ && \alpha\sum\limits_{m\ne n}
\frac{<m^{(0)}|\left[\hat\xi^4-2\left (\epsilon_m^{(0)}+\epsilon_n^{(0)}
\right)\hat\xi^2\right ]|n^{(0)}>}{n-m}
\,|m^{(0)}>.
\label{eq6-13}
\end{eqnarray}
Using the recurrence relation (\ref{eq6-10}), we get (note that $n!=\infty$ 
then $n<0$)
\begin{eqnarray} &&
\hat\xi^2|n^{(0)}>=\frac{1}{2}\left[\sqrt{\frac{(n+2)!}{n!}}\,|(n+2)^{(0)}>+\,
(2n+1)\,|n^{(0)}>+ \right . \nonumber \\ && \left .
\sqrt{\frac{n!}{(n-2)!}}\,|(n-2)^{(0)}>\right],
\label{eq6-14}
\end{eqnarray}
and
\begin{eqnarray} &&
\hat\xi^4|n^{(0)}>=\frac{1}{4}\left[\sqrt{\frac{(n+4)!}{n!}}\,|(n+4)^{(0)}>+
\,2(2n+3)\sqrt{\frac{(n+2)!}{n!}}\,|(n+2)^{(0)}>
\right . \nonumber \\ && \left . +3(2n^2+2n+1)\,|n^{(0)}>+
2(2n-1)\sqrt{\frac{n!}{(n-2)!}}\,|(n-2)^{(0)}>+
\right . \nonumber \\ && \left .
\sqrt{\frac{n!}{(n-4)!}}\,|(n-4)^{(0)}>\right].
\label{eq6-15}
\end{eqnarray}
Substituting these results into (\ref{eq6-13}), we obtain the final result
(which agrees with the previous works \cite{6-4,6-5})
\begin{eqnarray} &&
|n^{(1)}>=\frac{\alpha}{4}\left[-\frac{1}{4}\sqrt{\frac{(n+4)!}{n!}}\,
|(n+4)^{(0)}>+\,(2n+3)\sqrt{\frac{(n+2)!}{n!}}\,|(n+2)^{(0)}>
\right .\nonumber \\ && \left .
-(2n-1)\sqrt{\frac{n!}{(n-2)!}}\,|(n-2)^{(0)}>+
\frac{1}{4}\sqrt{\frac{n!}{(n-4)!}}\,|(n-4)^{(0)}>\right].
\label{eq6-16}
\end{eqnarray}
Let us return to the quantum Hamiltonian ${\cal H}={\cal H}_1-{\cal H}_2$.
From the above discussion it is clear that, to the first order in $\beta$,
its eigenvalues and eigenvectors have the form (in dimensionless variables)
\begin{eqnarray} &&
\epsilon_{n_1,n_2}=\epsilon_{n_1}-\epsilon_{n_2}=
(n_1-n_2)\left [1+\frac{3\alpha}{2}(n_1+n_2+1)\right],\;\;\;
\nonumber \\ &&
|n_1,n_2>=|n_1> \otimes \,|n_2>,
\label{eq6-17}
\end{eqnarray}
where $\epsilon_n=\epsilon_n^{(0)}+\epsilon_n^{(1)}$ and $|n>=|n^{(0)}>\,+\,
|n^{(1)}>$.
Note that the Planck corrections (a nonzero $\beta$) lift an 
$\infty$-order degeneracy inherent to the Koopman-von Neumann oscillator
\cite{6-6}. Maybe it is worthwhile to underline also \cite{6-6} that  
$\epsilon_{n_1,n_2}$ have nothing to do with the physical energy of the 
Koopman-von Neumann oscillator. They represent the possible eigenvalues of the 
evolution operator ${\cal{H}}$ which is not an observable in the 
Koopman-von Neumann theory.

The general time-dependent solution of the corresponding Schr\"{o}dinger 
equation is
\begin{equation}
\psi(\xi_1,\xi_2,t)=\sum\limits_{n_1,\,n_2=0}^\infty b_{n_1n_2}\,e^{-i\epsilon_
{n_1,n_2}\omega t}\,\psi_{n_1n_2}(\xi_1,\xi_2),
\label{eq6-18}
\end{equation}
where
\begin{eqnarray} &&
\psi_{n_1n_2}(\xi_1,\xi_2)=<\xi_1,\xi_2|n_1,n_2>=\psi_{n_1}(\xi_1)
\psi_{n_2}(\xi_2),\;\;\;
\nonumber \\ &&
\psi_n(\xi)=\psi_n^{(0)}(\xi)+\psi_n^{(1)}(\xi),
\label{eq6-19}
\end{eqnarray}
with $\psi_n^{(0)}(\xi)$ being defined by (\ref{eq6-8}), and according to 
(\ref{eq6-16})
\begin{eqnarray} &&
\psi_n^{(1)}(\xi)=\frac{\alpha}{4}\left[-\frac{1}{4}\sqrt{\frac{(n+4)!}{n!}}\,
\psi_{n+4}^{(0)}(\xi)+\,(2n+3)\sqrt{\frac{(n+2)!}{n!}}\,\psi_{n+2}^{(0)}(\xi)
\right .\nonumber \\ && \left .
-(2n-1)\sqrt{\frac{n!}{(n-2)!}}\,\psi_{n-2}^{(0)}(\xi)+
\frac{1}{4}\sqrt{\frac{n!}{(n-4)!}}\,\psi_{n-4}^{(0)}(\xi)\right].
\label{eq6-20}
\end{eqnarray}
The expansion coefficients $b_{n_1n_2}$ are determined by the initial wave
function $\psi(\xi_1,\xi_2,0)$:
\begin{equation}
b_{n_1n_2}=\int\limits_{-\infty}^\infty \int\limits_{-\infty}^\infty
\psi_{n_1n_2}(\xi_1,\xi_2)\,\psi(\xi_1,\xi_2,0)\,d\xi_1 d\xi_2.
\label{eq6-21}
\end{equation}
In the role of the initial wave function let's take
\begin{equation}
\psi(\xi_1,\xi_2,0)=\psi_{00}(\xi_1-\xi_0,\xi_2+\xi_0,0)=
\psi_0(\xi_1-\xi_0)\psi_0(\xi_2+\xi_0),
\label{eq6-22}
\end{equation}
where, according to (\ref{eq6-8}), (\ref{eq6-19}) and (\ref{eq6-20}),
\begin{equation} 
\psi_0(\xi\pm\xi_0)=\frac{e^{-\frac{(\xi\pm\xi_0)^2}{2}}}{\pi^{1/4}}\left [
1+\left .\left . \frac{\alpha}{64}\right (24\,H_2(\xi\pm\xi_0)-
H_4(\xi\pm\xi_0)\right )\right ].
\label{eq6-23}
\end{equation}
Then (\ref{eq6-21}) takes the form
\begin{equation}
b_{n_1n_2}=\int\limits_{-\infty}^\infty\psi_{n_1}(\xi_1)\psi_{0}(\xi_1-
\xi_0)d\xi_1\int\limits_{-\infty}^\infty\psi_{n_2}(\xi_2)\psi_{0}(\xi_2+
\xi_0)d\xi_2\equiv b_{n_1}(\xi_0)b_{n_2}(-\xi_0),
\label{eq6-24}
\end{equation}
where ($H_n(\xi)=0$ if $n<0$)
\begin{eqnarray} &&
\psi_{n}(\xi)=
\frac{e^{-\xi^2/2}}{\sqrt{2^nn!\sqrt{\pi}}}\left [H_n(\xi)+\frac{\alpha}{4}
\left (-\frac{1}{16}H_{n+4}(\xi)+\frac{2n+3}{2}H_{n+2}(\xi)-
\right .\right .\nonumber \\ && \left . \left .
\frac{2(2n-1)n!}{(n-2)!}H_{n-2}(\xi)+\frac{n!}{(n-4)!}H_{n-4}(\xi)
\right)\right].
\label{eq6-25}
\end{eqnarray}
Integrals needed in the calculation of $b_n(\xi_0)$ are of the form
\begin{eqnarray} &&
\int\limits_{-\infty}^\infty e^{-\frac{\xi^2+(\xi-\xi_0)^2}{2}}H_n(\xi)
H_m(\xi-\xi_0)\,d\xi= \nonumber \\ &&
\sqrt{\pi}e^{-\frac{\xi_0^2}{4}}\,\sum\limits_{k=0}^
{\min(n,m)}\frac{2^kn!m!}{k!(n-k)!(m-k)!}\xi_0^{n-k}\,(-\xi_0)^{m-k}.
\label{eq6-26}
\end{eqnarray}
This formula by itself can be proved by noting that
\begin{eqnarray} &&
\int\limits_{-\infty}^\infty e^{-\frac{\xi^2+(\xi-\xi_0)^2}{2}}H_n(\xi)
H_m(\xi-\xi_0)\,d\xi= \nonumber \\ &&
e^{-\frac{\xi_0^2}{4}}\int\limits_{-\infty}^\infty 
e^{-x^2}H_n\left(x+\frac{\xi_0}{2}\right)H_m\left(x-\frac{\xi_0}{2}
\right)\,dx,
\label{eq6-27A}
\end{eqnarray}
then using the addition theorem for Hermite polynomials \cite{6-7}
\begin{equation}
H_n(x+y)=\sum\limits_{k=0}^n\binom{n}{k}H_k(x)\,(2y)^{n-k},
\label{eq6-27}
\end{equation}
and the orthogonality property of the Hermite polynomials
\begin{equation}
\int\limits_{-\infty}^\infty e^{-x^2}H_n(x)H_m(x)=2^n n!\sqrt{\pi}\delta_{nm}.
\label{eq6-28}
\end{equation}
After a straightforward calculation and some algebra, we get the final result 
for $b_n(\xi_0)$:
\begin{eqnarray} &&
b_n(\xi_0)=\frac{e^{-\xi_0^2/4}\xi_0^n}{\sqrt{2^n n!}}\left[1+\frac{\alpha}{32}
\left(-\xi_0^4+12(n+2)\xi_0^2-12n(n+3)- 
\right . \right . \nonumber \\ && \left . \left .
16n(n-1)(n-2)\xi_0^{-2}\right)\right]
\equiv b_n^{(0)}(\xi_0)+\alpha b_n^{(1)}(\xi_0).
\label{eq6-29}
\end{eqnarray} 
Therefore the wave function takes the form
\begin{eqnarray} &&
\psi(\xi_1,\xi_2,t)=\psi_{(\xi_0)}(\xi,t)\psi_{(-\xi_0)}(\xi,t),\;\;\;
\\ &&
\psi_{(\xi_0)}(\xi,t)=\sum\limits_{n=0}^\infty b_n(\xi_0)e^{-in\omega t\left [
1+\frac{3\alpha}{2}(n+1)\right]}\psi_n(\xi)\approx \psi_{(\xi_0)}^{(c)}(\xi,t)+
\alpha \psi_{(\xi_0)}^{(nc)}(\xi,t), \nonumber
\label{eq6-30}
\end{eqnarray}
where
\begin{eqnarray} &&
\psi_{(\xi_0)}^{(c)}(\xi,t)=\sum\limits_{n=0}^\infty 
\frac{e^{-\xi_0^2/4}\xi_0^n}{\sqrt{2^n n!}}e^{-in\omega t
\left [1+\frac{3\alpha}{2}(n+1)\right]}\psi_n(\xi),
\nonumber \\ &&
\psi_{(\xi_0)}^{(nc)}(\xi,t)=\sum\limits_{n=0}^\infty b_n^{(1)}(\xi_0)
e^{-in\omega t}\psi_n^{(0)}(\xi).
\label{eq6-31}
\end{eqnarray}
Let's consider two parts of $\psi_{(\xi_0)}(\xi,t)$ separately. The first one
is a (generalized) coherent state. Namely, if we define generalized creation 
and annihilation operators \cite{6-4,6-5}
\begin{eqnarray} &&
\hat a=\hat a^{(0)}+\frac{\alpha}{4}\left (2\hat a^{(0)\,3}-6\hat N^{(0)}
\hat a^{(0)\,\dagger}+\hat a^{(0)\,\dagger \,3}\right ),\;\;\;
\nonumber \\ &&
\hat a^\dagger=\hat a^{(0)\,\dagger}+\frac{\alpha}{4}\left (2\hat a^{(0)\,
\dagger\,3}-6\hat a^{(0)}\hat N^{(0)}+\hat a^{(0)\,3}\right),
\label{eq6-32}
\end{eqnarray}
with $\hat N^{(0)}=\hat a^{(0)\,\dagger}\hat a^{(0)}$, then it can be checked 
by direct calculation that (up to the first order in $\alpha$)
\begin{equation}
\hat a |n>\,=\,\sqrt{n}\,|n-1>,\;\;\;
\hat a^\dagger |n>\,=\,\sqrt{n+1}\,|n+1>.
\label{eq6-33}
\end{equation}
Therefore we have the usual commutation relations
\begin{equation}
[\hat a,\,\hat a^\dagger]=1,\;\;\; [\hat N,\,\hat a]=-\hat a,\;\;\;
[\hat N,\,\hat a^\dagger]=\hat a^\dagger,\;\;\hat N=\hat a^\dagger\hat a.
\label{eq6-34}
\end{equation}
Using (\ref{eq6-33}), it can be easily checked that
\begin{equation}
\hat a |\xi_0(c)>=\frac{\xi_0}{\sqrt{2}}\,|\xi_0(c)>,\;\;\;
|\xi_0(c)>=\sum\limits_{n=0}^\infty \frac{e^{-\xi_0^2/4}\xi_0^n}
{\sqrt{2^n n!}}\,|n>.
\label{eq6-34A}
\end{equation}
Therefore $\psi_{(\xi_0)}^{(c)}(\xi,0)=<\xi|\xi_0(c)>$ is an eigenstate of 
the annihilation operator.

We can approximately solve (\ref{eq6-32}) and obtain
\begin{equation}
\hat a^{(0)}=\hat a-\frac{\alpha}{4}\left (2\hat a^3-6\hat N
\hat a^\dagger +\hat a^{\dagger \,3}\right ),\;\;\;
\hat a^{(0)\,\dagger}=\hat a^\dagger-\frac{\alpha}{4}\left (
2\hat a^{\dagger\,3}-6\hat a\hat N+\hat a^3\right),
\label{eq6-35}
\end{equation}
where $\hat N=\hat a^\dagger \hat a$. Then
\begin{equation}
\hat\xi=\frac{1}{\sqrt{2}}\left (\hat a^{(0)}+\hat a^{(0)\,\dagger}\right )=
\frac{1}{\sqrt{2}}\left [\hat a +\hat a^\dagger -\frac{3\alpha}{4}\left(
\hat a^3-2(\hat N\hat a^\dagger +\hat a\hat N)+\hat a^{\dagger\,3}\right) 
\right ],
\label{eq6-36}
\end{equation}
and
\begin{equation}
\hat\eta=\frac{1}{\sqrt{2}\, i}\left (\hat a^{(0)}-\hat a^{(0)\,\dagger}
\right )=\frac{i}{\sqrt{2}}\left [\hat a^\dagger -\hat a-\frac{\alpha}{4}\left(
\hat a^{\dagger\,3}+6(\hat N\hat a^\dagger -\hat a\hat N)-\hat a^3\right) 
\right ].
\label{eq6-37}
\end{equation}
From the commutation relations (\ref{eq6-34}), the validity of the following 
relations can be easily established:
\begin{eqnarray} &&
\hat a \hat N \hat a^\dagger=(\hat N +1)^2,\;\;\;\hat a^\dagger \hat N \hat a=
\hat N (\hat N -1),\;\;\; \hat a^2 \hat N=\hat a^2+\hat a \hat N \hat a,\;\;\;
\nonumber \\ &&
\hat a^{\dagger \,2} \hat N=-\hat a^{\dagger \,2}+\hat a^\dagger \hat N \hat 
a^\dagger, 
\hat a^2\hat a^{\dagger\,2}=(\hat N+1)(\hat N+2),\;\;\;
\nonumber \\ &&
\hat a \hat a^{\dagger\,3}=2\hat a^{\dagger \,2}+\hat a^\dagger \hat N \hat 
a^\dagger,\;\;\; \hat a^\dagger \hat a^3=-\hat a^2+\hat a \hat N \hat a.
\label{eq6-38}
\end{eqnarray}
Using these relations (and their Hermitian conjugates then needed), we obtain
\begin{eqnarray} &&
\xi^2\approx\frac{1}{2}\left [\hat a^2+\hat a^{+\,2}+2\hat N+1-
\frac{3\alpha}{4}\left (2\hat a^4+2\hat a^{+\,4}-\hat a(2\hat N+1)\hat a- 
\right . \right . \\ && \left . \left . \hat a^+
(2\hat N+1)\hat a^+ - 4(2\hat N^2+2\hat N+1)\right )\right ],\;\;\;
\eta^2\approx \frac{1}{2}\left [-\hat a^2-\hat a^{+\,2}+2\hat N+1+
\right . \nonumber \\ && \left .
\frac{\alpha}{4}\left (2\hat a^4+2\hat a^{+\,4}+5\hat a(2\hat N+1)\hat a+
5\hat a^+(2\hat N+1)\hat a^+ -12(2\hat N^2+2\hat N+1)\right )\right ],
\nonumber \\ &&
\alpha \eta^4\approx\frac{\alpha}{4}\left (\hat a^4+\hat a^{+\,4}
-2\hat a(2\hat N+1)\hat a-2\hat a^+(2\hat N+1)\hat a^+ +3(2\hat N^2+2\hat N+1)
\right ),
\nonumber
\label{eq6-39}
\end{eqnarray} 
and the Hamiltonian (\ref{eq6-6}) takes the form
\begin{equation}
\hat h=\hat N +\frac{1}{2}+\frac{3\alpha}{4}\left (2\hat N^2+2\hat N+1\right).
\label{eq6-40}
\end{equation}
Now we are ready to calculate the mean value
\begin{equation}
<\xi>\,\approx\, <\xi_0(c)|\hat \xi|\xi_0(c)>\,+\,\alpha\left (<\xi_0(c)|
\hat \xi |\xi_0(nc)>\,+\,<\xi_0(nc)|\hat \xi|\xi_0(c)>\right ).
\label{eq6-41}
\end{equation}
The first term conveniently can be calculated in the Heisenberg picture. The
time evolved annihilation operator is \cite{6-8}
\begin{eqnarray} &&
\hat a(t)=e^{i\omega t\hat h}\hat a(0)e^{-i\omega t\hat h}=
\\ && \hat a(0)+i\omega t
[\hat h,\,\hat a(0)]+\frac{(i\omega t)^2}{2!}[\hat h,\,[\hat h,\,\hat a(0)]]+
\frac{(i\omega t)^3}{3!}[\hat h,\,[\hat h,\,[\hat h,\,\hat a(0)]]]+\cdots
\nonumber
\label{eq6-42}
\end{eqnarray}
But from (\ref{eq6-40}) we have
\begin{eqnarray} &&
[\hat h,\,\hat a]=-\hat a\left(1+3\alpha\hat N\right )=
-\left[1+3\alpha\left(\hat N+1\right)\right ]\hat a,\;\;
\\ && 
[\hat h,\,[\hat h,\,\hat a]]=(-1)^2\,\hat a\left(1+3\alpha\hat N\right)^2=
(-1)^2\,\left[1+3\alpha\left(\hat N+1\right)\right]^2\hat a, \ldots
\nonumber
\label{eq6-43}
\end{eqnarray}
Therefore 
\begin{equation}
\hat a(t)=\hat a(0)\,e^{-i\omega t\left(1+3\alpha\hat N\right)}=
e^{-i\omega t\left[1+3\alpha\left(\hat N+1\right)\right]}\,\hat a(0).
\label{eq6-44}
\end{equation}
As we are interested in the classical motion, we can neglect the difference
between $\hat N$ and $\hat N+1$ operators in the above formulas and thus 
assume that $\hat a(0)$ and $e^{-i\omega t\left(1+3\alpha\hat N\right)}$, as 
well as $\hat a(0)$ (or $\hat a^\dagger(0)$) and $\hat N$ can be commuted
freely while calculating mean values (this is equivalent to assuming that 
$\xi_0^2\gg 1$). Besides, up to the first order in $\alpha$, we can change
$\hat N$ in the exponent $e^{-i\omega t\left(1+3\alpha\hat N\right)}$ by its
mean value $\xi_0^2/2$ in the coherent state $|\xi_0(c)>$. In light of (time
evolved) (\ref{eq6-36}), then we obtain
\begin{eqnarray} &&
<\xi_0(c)|\hat \xi|\xi_0(c)>= \\ &&
\xi_0\left(1+\frac{3\alpha}{4}\,\xi_0^2
\right )\,\cos{\left[\omega t\left(1+\frac{3\alpha}{2}\,\xi_0^2\right)\right]}
-\frac{3\alpha}{8}\,\xi_0^3\,\cos{\left[3\,\omega t\left(1+\frac{3
\alpha}{2}\,\xi_0^2\right)\right]}. \nonumber
\label{eq6-44A}
\end{eqnarray}

When calculating last two terms in (\ref{eq6-41}), we can assume $\alpha=0$
in the $\psi^{(c)}_{(\xi_0)}(\xi,t)$ wave function and get
\begin{equation}
\left .\psi^{(c)}_{(\xi_0)}(\xi,t)\right |_{\alpha=0}=\frac{e^{-(\xi_0^2-
\xi_0^2(t))/4}}{\pi^{1/4}}\,e^{-\frac{1}{2}\left(\xi-\xi_0(t)\right)^2},
\label{eq6-45}
\end{equation}
where $\xi_0(t)=\xi_0\,e^{-i\omega t}$. This expression for $\psi^{(c)}_
{(\xi_0)}(\xi,t)$ was obtained by using the following generating function for 
Hermite polynomials \cite{6-7}
\begin{equation}
F(t)\equiv e^{2xt-t^2}=\sum\limits_{n=0}^\infty H_n(x)\,\frac{t^n}{n!}.
\label{eq6-46}
\end{equation}
Differentiating this equation, we get respectively
\begin{eqnarray} &&
\sum\limits_{n=0}^\infty n\,H_n(x)\,\frac{t^n}{n!}=t\,\frac{dF(t)}{dt}=
2t(x-t)F(t),\;\;\;
\nonumber \\ &&
\sum\limits_{n=0}^\infty n(n-1)\,H_n(x)\,\frac{t^n}{n!}=t^2\,\frac{d^2F(t)}
{dt^2}=2t^2\left[2(x-t)^2-1\right ]F(t), \\ &&
\sum\limits_{n=0}^\infty n(n-1)(n-2)\,H_n(x)\,\frac{t^n}{n!}=t^3\,\frac{d^3
F(t)}{dt^3}=4t^3(x-t)\left[2(x-t)^2-3\right ]F(t). \nonumber
\label{eq6-47}
\end{eqnarray}
With the help of these relations, it is possible to calculate $\psi^{(nc)}_
{(\xi_0)}(\xi,t)$. The result is
\begin{equation}
\psi^{(nc)}_{(\xi_0)}(\xi,t)=\frac{e^{-(\xi_0^2-
\xi_0^2(t))/4}}{32\,\pi^{1/4}}\,e^{-\frac{1}{2}\left(\xi-\xi_0(t)\right)^2}
\left (A_1\xi^3+A_2\xi^2+A_3\xi+A_4\right ),
\label{eq6-48}
\end{equation}
where 
\begin{eqnarray} &&
A_1=-16\xi_0e^{-3i\omega t},\;\;\;\;A_2=12\xi_0^2e^{-2i\omega t}\left (
2e^{-2i\omega t}-1\right ), \nonumber \\ &&
A_3=12\xi_0e^{-i\omega t}\left [\xi_0^2\left(1+e^{-2i\omega t}-e^{-4i\omega t}
\right )+2\left (e^{-2i\omega t}-2\right)\right ], \nonumber \\ &&
A_4=-\xi_0^4\left (1+6e^{-2i\omega t}+3e^{-4i\omega t}-2e^{-6i\omega t}
\right )+  \nonumber \\ &&
2\xi_0^2\left (12+15e^{-2i\omega t}-6e^{-4i\omega t}\right ).
\label{eq6-49}
\end{eqnarray}
Now, using Gaussian integrals, it is straightforward to calculate
\begin{eqnarray} &&
<\xi_0(c)|\hat \xi |\xi_0(nc)>\,+\,\mathrm{c.c.}= 
\int\limits_{-\infty}^\infty d\xi 
\frac{e^{-(\xi-\xi_0\cos{\omega t})^2}}
{32\sqrt{\pi}}\,\xi\left [(A_1+A_1^*)\xi^3+
\right .\nonumber \\ && \left .
(A_2+A_2^*)\xi^2+(A_3+A_3^*) \xi+A_4+A_4^* \right ]. 
\label{eq6-50}
\end{eqnarray}
After some algebra and trigonometry, we will find rather surprisingly that 
almost everything cancels each other and the result takes a simple form
\begin{eqnarray} &&
\alpha \left (<\xi_0(c)|\hat \xi |\xi_0(nc)>\,+\,\mathrm{c.c.}\right )=
-\frac{3\alpha}{8}\xi_0(4+\xi_0^2)\cos{\omega t}\approx 
-\frac{3\alpha}{8}\xi_0^3\cos{\omega t}\approx
\nonumber \\ &&
-\frac{3\alpha}{8}\xi_0^3\,\cos{\left[\omega t\left(1+\frac{3\alpha}{2}\,
\xi_0^2\right)\right]}.
\label{eq6-50A}
\end{eqnarray} 
In combination with (\ref{eq6-44A}), this result implies
 \begin{eqnarray} &&
<\xi_1>=\xi_0\left(1+\frac{3\alpha}{8}\,\xi_0^2
\right )\,\cos{\left[\omega t\left(1+\frac{3\alpha}{2}\,\xi_0^2\right)\right]}
\nonumber \\ &&
-\frac{3\alpha}{8}\,\xi_0^3\,\cos{\left[3\,\omega t\left(1+\frac{3
\alpha}{2}\,\xi_0^2\right)\right]}.
\label{eq6-44B}
\end{eqnarray}
To get $<\xi_2>$, we should make changes $\xi_0\to -\xi_0$ and $\omega\to -
\omega$ into this expression and thus  $<\xi_2>\,=\,-<\xi_1>$. Then 
(\ref{eq6-1}) and (\ref{eq6-7}) indicate that $<Q>=0$, while
\begin{eqnarray} &&
<q>=\sqrt{\frac{2\hbar}{m\omega}}\,\xi_0\left\{
\left(1+\frac{3\alpha}{8}\,\xi_0^2
\right )\,\cos{\left[\omega t\left(1+\frac{3\alpha}{2}\,\xi_0^2\right)\right]}
\right . \nonumber \\ && \left .
-\frac{3\alpha}{8}\,\xi_0^2\,\cos{\left[3\,\omega t\left(1+\frac{3
\alpha}{2}\,\xi_0^2\right)\right]}\right \}.
\label{eq6-51}
\end{eqnarray}
As we see, for the initial state considered, the mean value of the hidden 
variable $Q$ remains strictly zero (up to the first order in $\alpha$), while
the effect of the Planck scale physics on the mean value of the classical
variable $q$ is twofold. Namely, a small admixture of the third-harmonic 
appears in the time evolution of the classical oscillator and its period of
oscillations is modified according to
\begin{equation}
T=\frac{2\pi}{\omega}\,\left (1-\frac{3\alpha}{2}\,\xi_0^2\right).
\label{eq6-52}
\end{equation}    
In fact (\ref{eq6-52}) is the same result as we have obtained earlier in 
(\ref{eq4-34}). Indeed, $p_0$ in (\ref{eq4-34}) is the maximum momentum $p_m$
of the oscillations. In the case of (\ref{eq6-51}), the amplitude of the 
oscillations is $q_m\approx \sqrt{\frac{2\hbar}{m\omega}}\,\xi_0$ which 
corresponds to the energy $E=\frac{1}{2}\,m\omega^2q_m^2\approx \hbar\omega
\xi_0^2$ (thus the condition of classicality $\xi_0^2\gg 1$ is the same as
$E\gg\hbar\omega$), and 
\begin{equation}
p_m^2=2mE=2m\hbar\omega\xi_0^2=\frac{2\alpha}{\beta}\,\xi_0^2,
\label{eq6-53}
\end{equation}  
which proves the equivalence of (\ref{eq4-34}) and (\ref{eq6-52}).

\section{Concluding remarks}
In this note we have tried to combine Koopman-von Neumann-Sudarshan 
perspective on classical mechanics with the generalized uncertainty principle.
We have considered two versions of the generalized commutation relations.
The results were similar: classical position and momentum operators cease to 
be commuting and hidden variables show themselves explicitly in classical 
evolution equations. In situations then the effect of these hidden variables
can be neglected in evolution equations, the modification of classical dynamics
is similar (but not identical) to the modification obtained by using more 
traditional approach of replacement of commutators by Poisson brackets.

We suspect that the above mentioned features are common for a large class of
generalized uncertainty principle based models if they are interpreted in the
Koopman-von Neumann-Sudarshan framework. Therefore, from this perspective,
we can conclude that Planck scale quantum gravity effects destroy classicality.
However this breakdown of classicality is controlled by a small dynamical 
parameter $\frac{p^2}{p_P^2}$ and can be neglected for all practical purposes 
thanks to the huge hierarchy between the masses of ordinary particles and the 
Planck mass $m_P=1.2\times 10^{19}~\mathrm{GeV}/c^2$. Usually this huge 
hierarchy is considered as a problem to be explained \cite{5-1}. As we see, 
for classicality it can be beneficial. For a macroscopic body the effective 
deformation parameter $\beta$ is approximately $N^2$ times smaller than for 
its elementary constituents, where $N$ is the number of constituents 
\cite{4-8}. Therefore macroscopic bodies, notwithstanding their large momenta,
provide no advantage in observing Planck scale induced non-classical effects,
as the small parameter controlling these non-classical effects for macroscopic
bodies  becomes $\frac{p^2}{N^2 p_P^2}$.

It should be noted that when a modified quantum-mechanical commutator is 
considered, a self-consistent description in terms of macroscopic center-of-mass 
coordinates is not straightforward. If the equations of motions that follow from
the modified commutation relations are naively applied to the macroscopic 
objects like planets in the solar system, then an unacceptably large corretions to 
their classical dynamics will follow, unless the deformation parameter is unreasonably
small. This problem is common to many quantum-gravity inspired deformations of 
quantum mechanics and is known as the soccer-ball problem \cite{5-1A}.

Despite various attempts to address the soccer-ball problem, so far none has been 
generally accepted. As was already mentioned, it was argued that the deformation
parameter is expected to decrease with the number of constituent particles in the
macroscopic object \cite{4-8} (see also \cite{5-1B}). There are other approaches to 
the classical limit of the deformed quantum mechanics which do not assume such
scaling of the deformation parameter with the number of constituent particles 
(see, for example, \cite{5-1C,5-1D} and references therein).

Let us emphasize that our main result, that the Planck-scale quantum gravity effects
destroy classicality, if classicality is understood from the Sudarshan's perspective
on the Koopman-von Neumann mechanics, does not depend on the specific
solution of the soccer-ball problem. From this point of view, classical dynamics is not 
just a limiting case of quantum dynamics, but dynamics that is realized in a part of a 
quantum system due to a very special quantum dynamics that affects this quantum 
system. At that some quantum variables of the encompassing quantum system
remain hidden for classical observers. Quantum gravity effects, due to a universal
character of gravity, is expected to deform quantum evolution of all degrees of
freedom of the encompassing quantum system. As a result, the classical subsystem
will be no longer decoupled from the hidden variables and the classicality will be
destroyed. Although we have demonstrated this feature only for some specific 
realizations of deformed quantum mechanics with generalized commutation relations, 
we expect that this destruction of classicality, when the quantum dynamics of  the 
encompassing system is deformed due to  the Planck-scale quantum gravity effects,
to be a fairly general result.

It may happen that the interrelations between quantum mechanics, classical
mechanics and gravity are much more tight and intimate than anticipated. 
The imprints left by quantum mechanics in classical mechanics are more 
numerous than is usually believed \cite{5-2,5-3}. In fact the mathematical 
structure that allows quantum mechanics to emerge already exists in classical 
mechanics \cite{5-4}. Particularly surprising, maybe, is that 
Schr\"{o}dinger-Robertson uncertainty principle has an exact counterpart in
classical mechanics which can be formulated using some subtle developments
in symplectic topology, namely Gromov's non-squeezing theorem and the related 
notion of symplectic capacity \cite{5-5}.

On the other hand there are unexpected and deep relations between gravity and
quantum mechanics, in particular between Einstein-Rosen wormholes and quantum  
entanglement \cite{5-6,5-7}.

We believe the Koopman-von Neumann formulation of classical mechanics might
be useful in investigating a twilight zone between quantum and classical
mechanics. ``It deserves to be better known among physicists, because it gives 
a new perspective on the conceptual foundations of quantum theory, and it may 
suggest new kinds of approximations and even new kinds of theories'' 
\cite{3-13A}. 

\section*{Acknowledgments}
We thank Carlo Beenakker for indicating useful references and Pasquale Bosso
for helpful correspondence. We are grateful to our colleagues for their
comments that helped to improve the manuscript. The work of Z.K.S. is 
supported by the Ministry of Education and Science of the Russian Federation.

\end{document}